\documentclass[twocolumn,showpacs,preprintnumbers,superscriptaddress,amsmath,amssymb,floats]{revtex4}

\usepackage{graphicx}
\usepackage{dcolumn}
\usepackage{bm}
\usepackage{longtable}

\begin{document}

\title{Experimental study of fragmentation products in the reactions $^{112}$Sn + $^{112}$Sn \\
and $^{124}$Sn + $^{124}$Sn at 1$\cdot A$ GeV}

\author{V. F\"ohr\footnote{This work forms part of the PhD thesis of V. F\"ohr.}}
 \email{ville.t.fohr@jyu.fi}
\affiliation{%
GSI---Helmholtzzentrum f\"ur Schwerionenforschung GmbH., D-64291 Darmstadt, Germany}%
\affiliation{%
Department of Physics, P.O. Box 35 (YFL) FI-40014 University of Jyv\"askyl\"a, Finland.}%

\author{A. Bacquias}%
\affiliation{%
GSI---Helmholtzzentrum f\"ur Schwerionenforschung GmbH., D-64291 Darmstadt, Germany}%

\author{E. Casarejos\footnote{Present address: Univ. of Vigo, E-36310 Vigo, Spain}}
\affiliation{%
Universidad de Santiago de Compostela, E-15706, Spain.}%

\author{T. Enqvist\footnote{Present address: Oulu Southern Institute and Department of Physics, University of Oulu, Finland}}
\affiliation{%
CUPP Project, P.O. Box 22, FI-86801, Pyh\"asalmi, Finland}%

\author{A. R. Junghans}
\affiliation{%
Helmholtz-Zentrum Dresden-Rossendorf, D-01314 Dresden, Germany}%

\author{A. Keli\ifmmode \acute{c}\else \'{c}\fi{}-Heil}
\affiliation{%
GSI---Helmholtzzentrum f\"ur Schwerionenforschung GmbH., D-64291 Darmstadt, Germany}%

\author{T. Kurtukian}
\affiliation{%
Universidad de Santiago de Compostela, E-15706, Spain.}%

\author{S. Luki\ifmmode \acute{c}\else \'{c}\fi{}\footnote{Present address: Vin\ifmmode \acute{c}\else \v{c}\fi{}a Institute of Nuclear Sciences}}
\affiliation{%
GSI---Helmholtzzentrum f\"ur Schwerionenforschung GmbH., D-64291 Darmstadt, Germany}%

\author{D. P\ifmmode \acute{e}\else \'{e}\fi{}rez-Loureiro}
\affiliation{%
Universidad de Santiago de Compostela, E-15706, Spain.}%

\author{R. Pleska\ifmmode \acute{c}\else \v{c}\fi{}}
\affiliation{%
GSI---Helmholtzzentrum f\"ur Schwerionenforschung GmbH., D-64291 Darmstadt, Germany}%

\author{M. V. Ricciardi}
\affiliation{%
GSI---Helmholtzzentrum f\"ur Schwerionenforschung GmbH., D-64291 Darmstadt, Germany}%

\author{K.-H. Schmidt}
\affiliation{%
GSI---Helmholtzzentrum f\"ur Schwerionenforschung GmbH., D-64291 Darmstadt, Germany}%

\author{J. Ta\ifmmode \ddot{i}\else \"{i}\fi{}eb}
\affiliation{CEA/DAM \ifmmode \acute{I}\else \^{I}\fi{}le-de-France, Bruy\ifmmode \breve{e}\else \`{e}\fi{}res-le-Ch\ifmmode \acute{a}\else \^{a}\fi{}tel, 91297 Arpajon, France}%

\date{\today}

\begin{abstract}
Production cross-sections and longitudinal velocity distributions
of the projectile-like residues produced in the reactions
$^{112}$Sn + $^{112}$Sn and $^{124}$Sn + $^{124}$Sn both at an
incident beam energy of 1$\cdot A$ GeV were measured with the
high-resolution magnetic spectrometer, the Fragment Separator
(FRS) of GSI. For both reactions the characteristics of the
velocity distributions and nuclide production cross sections were
determined for residues with atomic number $Z$ $\geq$ 10. A comparison 
of the results of the two reactions is presented.
\end{abstract}


\maketitle

\section{\label{sec:Intro}introduction}
Since heavy-ion beams at relativistic energies $E$ $>$100$\cdot A$ MeV became available
in laboratories \cite{PhysRevLett.52.180,Kienle1988277}, a possibility to study
static and dynamic properties of nuclear matter over a wide range
of temperature and density has been opened
\cite{annurev.ns.28.120178,Pochodzalla:1997vq}. Depending on the
impact parameter $b$, heavy-ion collisions can be divided into
three groups \cite{PhysRev.75.1779}:

One extreme are central collisions in which projectile and target
completely overlap. In this type of collisions, high densities and
high excitation energies can be achieved \cite{Bertsch1988189},
and thus they appear to be an excellent tool to study the equation
of state of hot and compressed nuclear matter as well as in-medium
nucleon-nucleon interactions. To this goal, immense experimental
effort has been, and is still being, invested to measure for
example the flow pattern of nucleons and particles,  kaon
production or charged-particles correlation in central heavy-ion
collisions \cite{Reisdorf:1997fx,Senger:2004xa,Reisdorf2010366}.
Since high densities and high excitation energies are achieved
only for short time intervals of the order $10^{-22}$ s and in
volumes of the order 100 fm$^3$ \cite{Cubero1990345}, it is mandatory 
to understand the complete dynamic evolution of the reaction in order to
extract the information on the nuclear equation of state under these extreme
conditions. This is still not an easy task.

An other extreme is the case of large impact parameters leading to very
peripheral collisions. This type of collisions is
characterized by a small mass loss in projectile and/or target and
rather low excitation energies. Projectile-like fragments move
with velocities very close to the original one of the projectile.
These collisions have been proved to be an excellent tool to study
e.g. nuclear-structure effects at large deformations
\cite{Schmidt2000221,Steinhauser199889,Rejmund2000215} or neutron
skin \cite{PhysRevC.76.051603}.

For the intermediate range of impact parameters, a considerable
amount of excitation energy \cite{Schmidt1993313} and a slight
linear momentum transfer are induced, but compression is small. Thus, the
mid-peripheral heavy-ion collisions at relativistic energies are
an ideal scenario for studying multifragment decay of the
spectator matter due to purely thermal instabilities
\cite{Schuttauf1996457}, avoiding any compression effect.
Multifragmentation reactions have been extensively studied in
order to search for the signals of the liquid-gas phase transition
in finite nuclear systems
\cite{Pochodzalla:1997vq,PhysRevLett.75.1040,Trautmann2005407}.
Since some time, isotopic effects in multifragmentation reactions
also gained a lot of interest
\cite{PhysRevC.65.044610,PhysRevLett.102.122701,PhysRevLett.102.142503,trautmann2008},
as neutron-star models or supernova simulations demand a nuclear
equation of state similar to those met in mid-peripheral relativistic heavy-ion collisions
\cite{PhysRevLett.72.2835,Dean1995429,Botvina2004233}. Similar to
the experiments where central collisions are studied, a lot of
effort is invested in developing devices covering the full solid
angle in order to attain particle multiplicities as well as
correlations between observed particles.

Recently, high-resolution experiments on kinematical properties of
projectile residues produced in mid-peripheral heavy-ion
collisions have been proposed as a new tool to study the nonlocal
properties of the nuclear force
\cite{PhysRevC.64.034601,PhysRevLett.90.212302}. According to the
model calculations \cite{PhysRevC.64.034601}, the transversal and
the longitudinal momentum distributions of the spectator matter
surviving the collisional stage are influenced by the participant
blast, occurring after the compression phase in the colliding
zone. Consequently, the momentum distributions of spectator
residues in mid-peripheral collisions should be sensitive to the
nuclear force. In order to yield conclusive results, 
the momentum distributions of projectile residues have
to be measured with high precision. This can only be achieved with high-resolution magnetic
spectrometers, as experimental set-ups covering full solid
angle do not have the required resolution.

Unfortunately, detailed experimental information on kinematical
properties of projectile residues produced in heavy-ion collisions
at relativistic energies is rather lacking. In a review on
measured mean velocities of spectator-like fragments presented by
Morrissey in 1989 \cite{PhysRevC.39.460} a clear correlation
between the observed momentum shift with the mass loss in very
peripheral collisions has been observed. This shift has been
interpreted as the consequence of  friction in the
nucleus-nucleus collision \cite{PhysRevC.39.460,AbulMagd1976327}.
On the other hand, the momentum distributions of lighter
fragments, with a mass loss larger than about one-third of the
mass of the projectile, respectively, the target nucleus, showed a
large spreading with no clear tendency. Since then, a lot of new data on
the momentum distributions have been measured, but unfortunately
only few of them cover the whole range - from projectile down to
the lowest nuclear charges of produced fragments \cite{Weber1994659,PhysRevC.70.054607,PhysRevC.76.064609,PhysRevC.78.044616}. To overcome this lack of high-precision data on the velocities of projectile fragments, a dedicated experimental campaign
\cite{PhysRevLett.90.212302,vladimir} has been started at GSI
using the heavy-ion accelerator SIS18 and the Fragment Separator (FRS).

The present work represents the next step in this campaign, and is
dedicated to a study of the influence of the isotopic composition of
the projectile on the kinematical properties of projectile residues in
peripheral and mid-peripheral relativistic heavy-ion collisions.
To this goal, two symmetric systems $^{112}$Sn+$^{112}$Sn and
$^{124}$Sn+$^{124}$Sn at the projectile energy of 1$\cdot A$ GeV
have been studied. The $N$/$Z$ ratio of $^{112}$Sn is 1.24, and
the one of $^{124}$Sn is 1.48, resulting, for a given $Z$, in the largest span in $N$/$Z$ values for stable nuclei in this mass range. For this exploratory study the beams of
stable nuclei have been chosen as their emittance is smaller than
in case of secondary beams, while available intensities are
higher. Since in both reactions the target and projectile are the same nuclei, the $N$/$Z$ stays homogeneous for all possible impact parameters, despite the small effects coming from the neutron skin. This $N$/$Z$ value is determined entirely by the corresponding tin nuclei in the system. The incident energy is chosen in such way to 
have the best conditions for the transmission of the reaction products through the FRS. 

In addition to the high-precision data on the longitudinal velocity
of the projectile fragments, also the production cross sections
have been measured.

The present work is ordered in the following way: In section II we
describe the experimental approach and the data analysis. In section III
velocity distribution of the final residues, as well as the
moments of this distribution are presented. In section IV
production cross sections of the projectile residues measured in
these two reactions are given. Detailed discussion on the physics
behind the data as well as comparison with different theoretical
predictions is a topic of forthcoming publications, and will not
be discussed here.

\section{\label{sec:experiment}Experiment and data analysis}
In this section details on the experimental set-up as well as
steps needed to be undertaken to obtain the velocity distributions
and production cross sections of all the projectile-like residues
will be presented.
\subsection{\label{sec:technique}Experimental technique}
\begin{figure}
\includegraphics[width=0.5\textwidth]{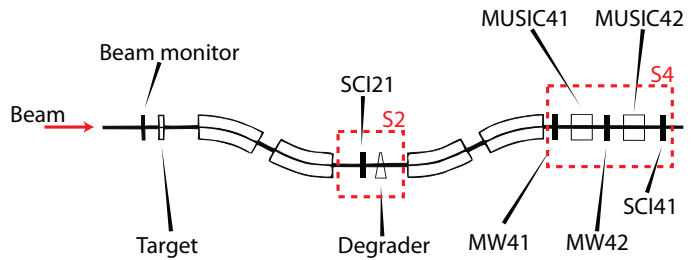}
\caption{\label{fig:setup}(Color online) Schematic view of the Fragment Separator (FRS) and the associated detector equipments. The notation of different detectors is explained in the text.}
\end{figure}

The experiment was performed at GSI,
Darmstadt, with two systems: $^{112}$Sn + $^{112}$Sn and
$^{124}$Sn + $^{124}$Sn both at an incident beam energy of 1$\cdot
A$ GeV. Beams were delivered from the universal linear accelerator
(UNILAC) to the SIS18 heavy-ion synchrotron where they were
extracted and guided through the target area to the FRagment
Separator (FRS) \cite{Geissel:1991zn}. The FRS is a
two-stage magnetic spectrometer with a maximum bending power of 18
Tm, an angular acceptance of 15 mrad around the beam axis, and a
momentum acceptance of 3\%. The FRS was used for the separation
and analysis of the reaction products. In Fig. \ref{fig:setup} a
schematic view of the experimental setup is shown with all the
detectors used in the experiment. The two tin beams impinged on
the tin targets whose isotopic composition closely corresponded to
the nuclei of the beam:  a 126.7 $\pm$ 0.6 mg/cm$^{2}$ thick
$^{112}$Sn target with 99.5$\pm$0.2 enrichment and a 141.8 $\pm$
0.7 mg/cm$^{2}$ thick $^{124}$Sn target with 97.5$\pm$0.2
enrichment. Due to the high linear momenta of the incoming projectiles, most of the produced
projectile-like fragments escaped the target in forward direction
and were then analyzed by the fragment separator FRS, used as a
momentum-loss achromat.
The scintillation detectors were used to acquire the horizontal position of
the passing ions and to register the start and the stop time
signals for the time-of-flight ({\it ToF}) measurement. The uncertainty
in the position determination was about 3 mm (FWHM), and in {\it ToF} it
amounted to 100 ps (FWHM). One scintillation detector was placed at the end
of the first stage – at the intermediate focal plane S2 – and
another one at the end of the second stage – at the final focal
plane S4.

The radius of the fragment trajectory $\rho $ is measured with a relative uncertainty of
about $4 \cdot 10^{-4}$. The magnetic field strength $B $ is
measured with high precision ($\Delta B/B \sim 10^{-4}$) using the
Hall probes. In this way, one can obtain a measure of the magnetic
rigidity $B\rho$ with a resolution of about 5$\cdot$10$^{-4}$.
Together with the longitudinal velocity, determined from the {\it ToF}
measurement, the mass-over-charge ratio $A$/$Z$ of the fragments
could be determined according to the formula:
\begin{equation}
\frac{A}{Z} = \frac{1}{c}\cdot \frac{e}{m_0+\delta m} \cdot \frac{B\rho}{\beta \gamma},
\label{eq:azidentification}
\end{equation}
where $c$ is the velocity of light, $e$ the elementary charge,
$m_0$ atomic mass unit, $\delta m = dM/A$ the mass excess per
nucleon, $\gamma = \sqrt{(1-\beta^2)^{-1}}$ the Lorentz factor,
$\beta =v/c$ velocity in natural units, where $v$ is the
longitudinal velocity of the fragment obtained from the {\it ToF}
measurement. For the calculation of the mass excess a generalized empirical
mass formula was used \cite{Myers19661}, which
provided sufficient accuracy for the $A/Z$ calculation. 

\begin{figure}
\includegraphics[width=0.5\textwidth]{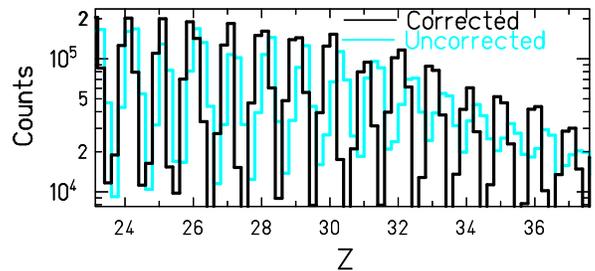}
\caption{\label{fig:muscide}(Color online) MUSIC charge resolution before (light blue) and after various corrections described in the text (black).}
\end{figure}

Due to their high velocity, the fragments were completely stripped
of electrons with a probability higher than 99\% \cite{Stohlker1991408}, so
that the charge of the passing ion $Q$ coincides with the atomic
number of the fragment $Z$. At the end of the second stage the
ions were detected by two multiple-sampling ionization chambers
(MUSICs) \cite{Pfutzner1994213}. The MUSICs provided the energy-loss signals which were
used to obtain the information on nuclear charge $Z$. Drift-time signals from the two MUSIC detectors also provided information on the horizontal position and the horizontal angle of the passing ions trajectory. This information was used to determine the length of the ions path between the scintillators which then allowed the determination of the velocity together with the {\it ToF} information. The nuclear-charge resolution has been improved by correcting the
energy-loss signal for the position and velocity dependence as discussed in e.g. ref. \cite{PhysRevC.78.044616}. In Fig. \ref{fig:muscide}, the Z spectrum deduced from the MUSIC 
energy-loss is shown before and after these corrections. The improvement in the
resolution is especially seen for the higher nuclear charges. The
atomic number $Z$ was determined with an uncertainty of $\Delta Z=0.4$ units (FWHM). In
order to prevent the overloading of the MUSIC detectors with
fragments produced with high counting rates i.e. light particles, the
threshold on the signals collected from the MUSICs was set so that the fragments with nuclear charge $Z \geq$ 10 were fully recorded. 
Due to the limited momentum acceptance of $\Delta p/p \cong \pm$ 1.5\% of the fragment separator
several magnetic settings were required to scan the full momentum distributions of the projectile
residues. The term "magnetic setting" refers to a measurement
performed with given magnetic-field values in the magnets.
The setting thus determines the acceptance of particles, with certain
range of magnetic rigidities, that are able to pass the FRS.

In some settings an additional degrader is added in between the
two stages of the FRS. This is beneficial when measuring fragments
produced with low yields. In these settings one can increase the
intensity of the beam without overloading the detectors with
fragments produced with higher counting rates and thus obtain a
proper statistics for all the fragments. In both systems the
settings that were devoted to measure the lighter residues, $Z$
$\leq$ 35, where measured with the aluminum degrader with a
thickness of (737.1 $\pm$ 1) mg$\cdot$ cm$^{-2}$.

\begin{figure}
\begin{center}
$\begin{array}{c}
\mbox{\bf $^{112}$Sn + $^{112}$Sn} \\
\includegraphics[width=0.5\textwidth]{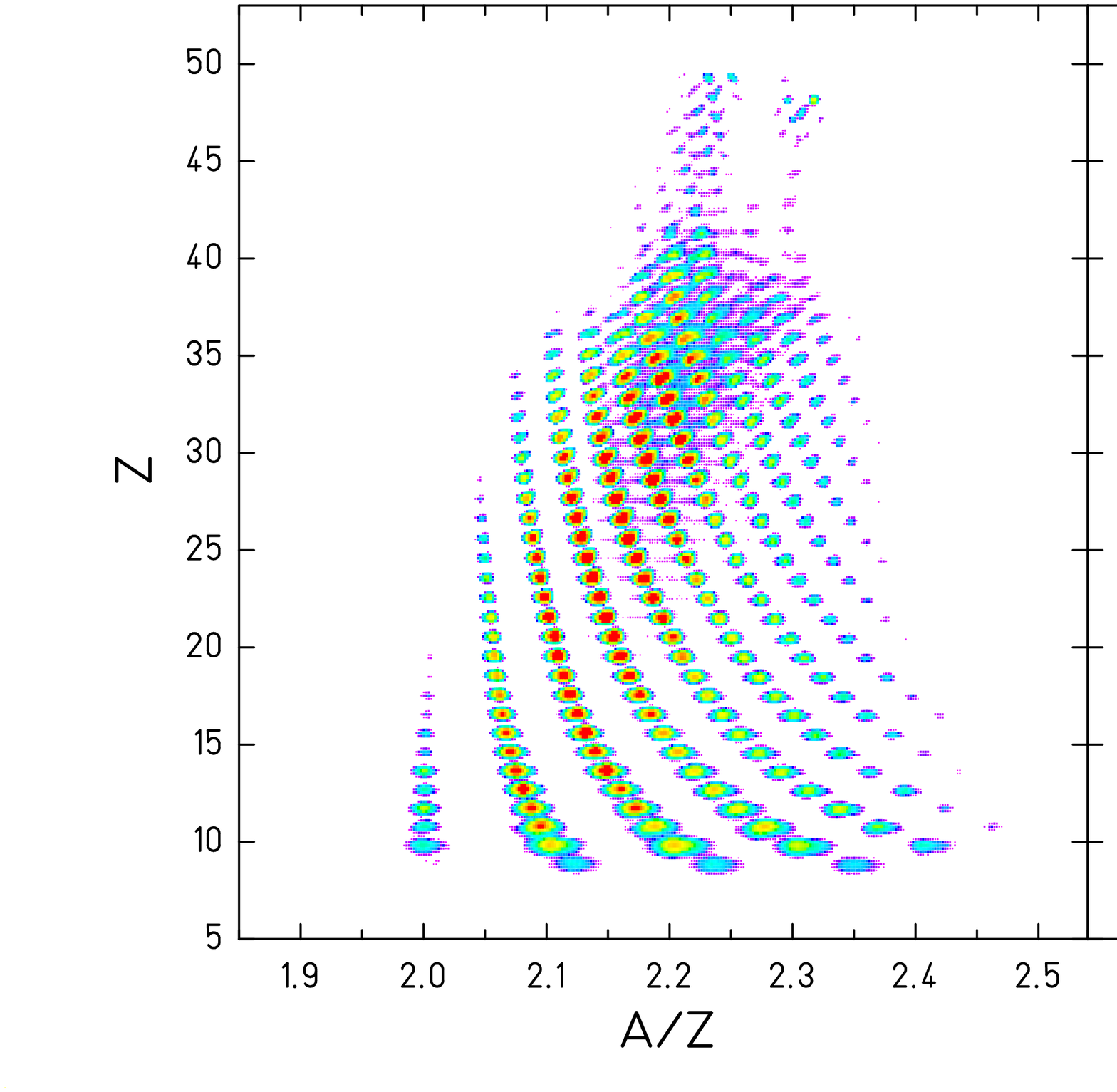}\\
\mbox{\bf $^{124}$Sn + $^{124}$Sn} \\
\includegraphics[width=0.5\textwidth]{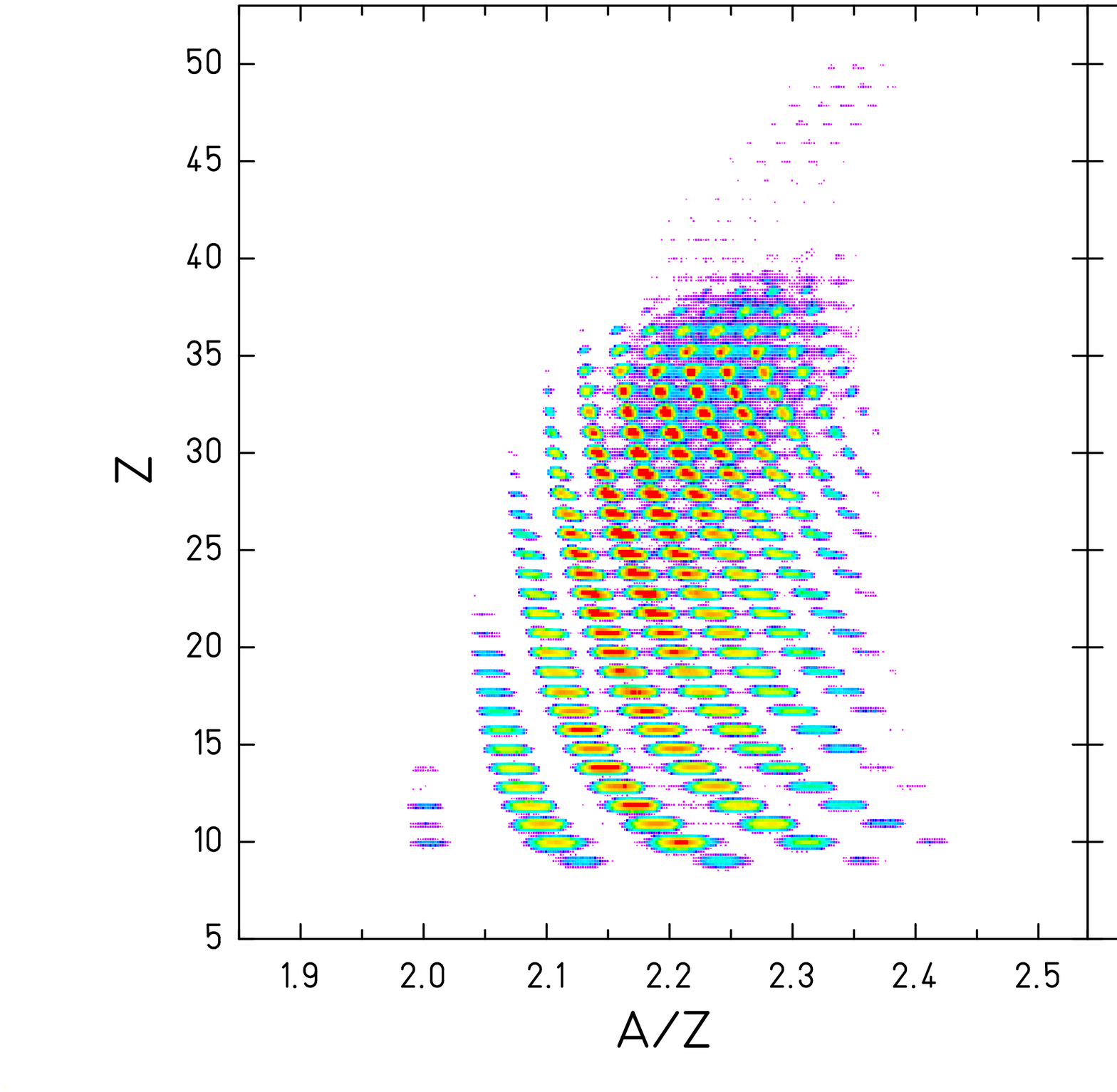}\\
\end{array}$
\end{center}
\caption{\label{fig:recog}(Color online) Recognition patterns of fragments formed in $^{112}$Sn +
$^{112}$Sn (top) and in $^{124}$Sn + $^{124}$Sn (bottom) reactions. Settings
measured with degrader (left) are shown separately from those
obtained without degrader (right). Color code represents yields on the logarithmic scale.}
\end{figure}

The recognition pattern, formed by plotting the $A$/$Z$ ratio
versus the charge of the fragments produced in both experiments
is presented in Fig. \ref{fig:recog}. Settings dedicated for
the measurement of light residues are given separately from the
heavy-residue recognition plots. The gaps in the recognition pattern appear due to
the necessity to protect the detectors against the primary beam. Therefore, a number of nuclei with magnetic rigidities very close to the beam were not measured. Each spot in Fig. \ref{fig:recog}
represents one nucleus with a given $A$ and $Z$. Using the
characteristic pattern for $N=Z$ nuclei, i.e. vertical line at
$A/Z=2$ and also the measurements with the primary beam, the full
identification of all the residues has been performed. The achieved
resolution in mass was $\Delta A/A=4\cdot 10^{-3}$.  

Once the fragments were isotopically identified in nuclear mass
$A$ and nuclear charge $Z$, equation \ref{eq:azidentification}
could be used to extract the longitudinal velocity from the known
$B \rho$ values at S2. In this way, the resolution in the
longitudinal velocity is given only by the resolution in the magnetic rigidity, 
as $A$ and $Z$ are integer numbers and thus contain no uncertainty, and amounts to $\Delta v/v = 10^{-4}$ representing about one order of magnitude improvement relative to the resolution obtained via {\it ToF} measurement. Thus, in the following, we will use the velocity
distributions obtained via "$B\rho$ measurement".
\subsection{Data analysis}
\subsubsection{Reconstruction of full velocity distributions}
As mentioned earlier, the FRS momentum acceptance in each setting
is $\pm$ 1.5\% of the magnetic rigidity ($B\rho$) of the selected
central trajectory. For each setting, the magnetic fields of
the dipoles were scaled by steps of 1.5\% to assure a sufficient
overlap of the velocity distributions measured in the neighboring
settings. Especially for lighter nuclei ($Z$ $<$ 30), the velocity
distributions are generally always wider then what one can measure
in one setting, and it is necessary to combine data from several
measurements with different magnetic rigidities to cover the full
range of velocities of each fragment.

Data obtained from different settings have to be normalized to the
primary-beam intensity, and corrected for the limited angular
acceptance and dead-time of the data-acquisition system before
being merged. These different corrections are described below.

\begin{figure}
\begin{center}
$\begin{array}{c}
\includegraphics[width=0.45\textwidth]{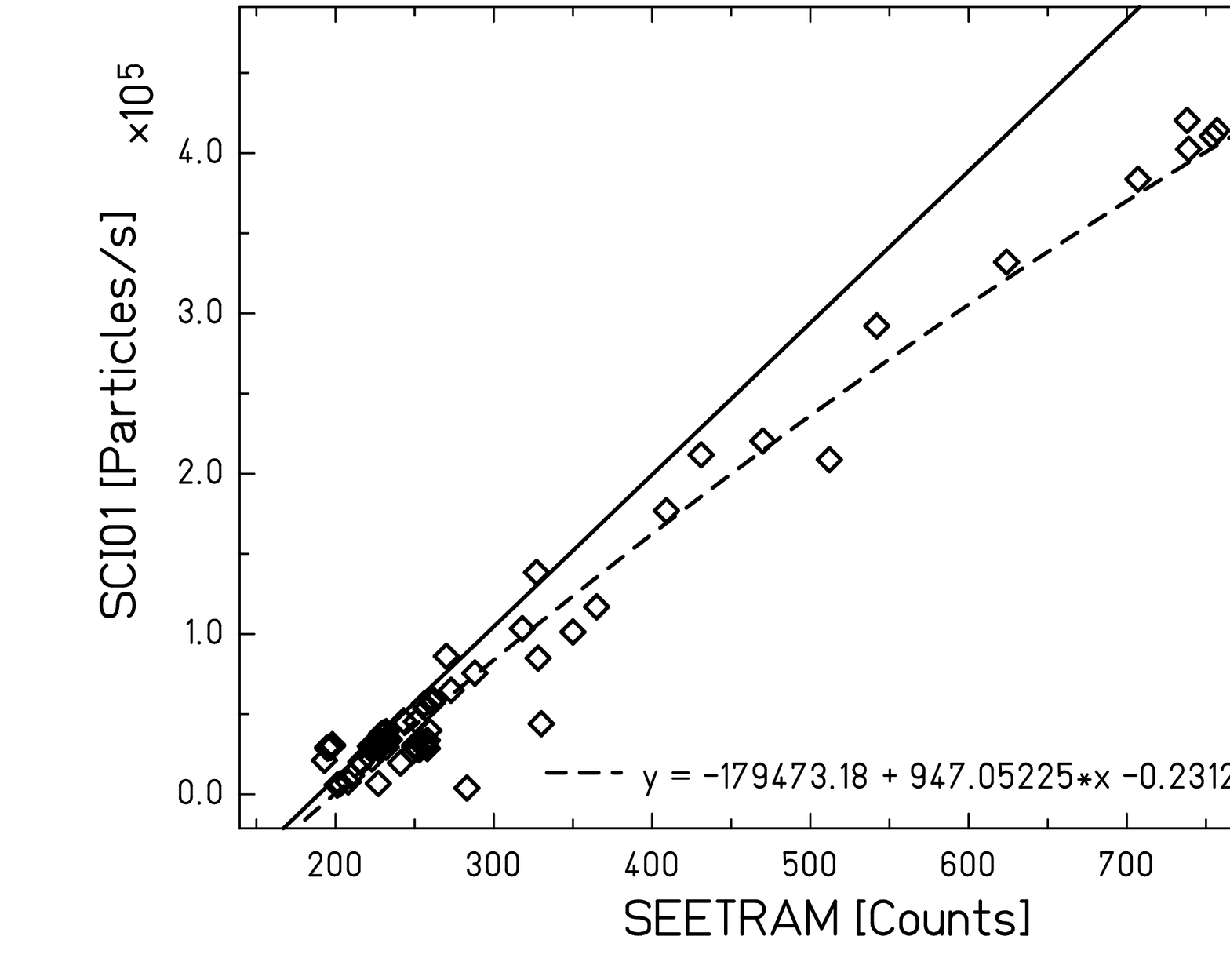}\\
\includegraphics[width=0.45\textwidth]{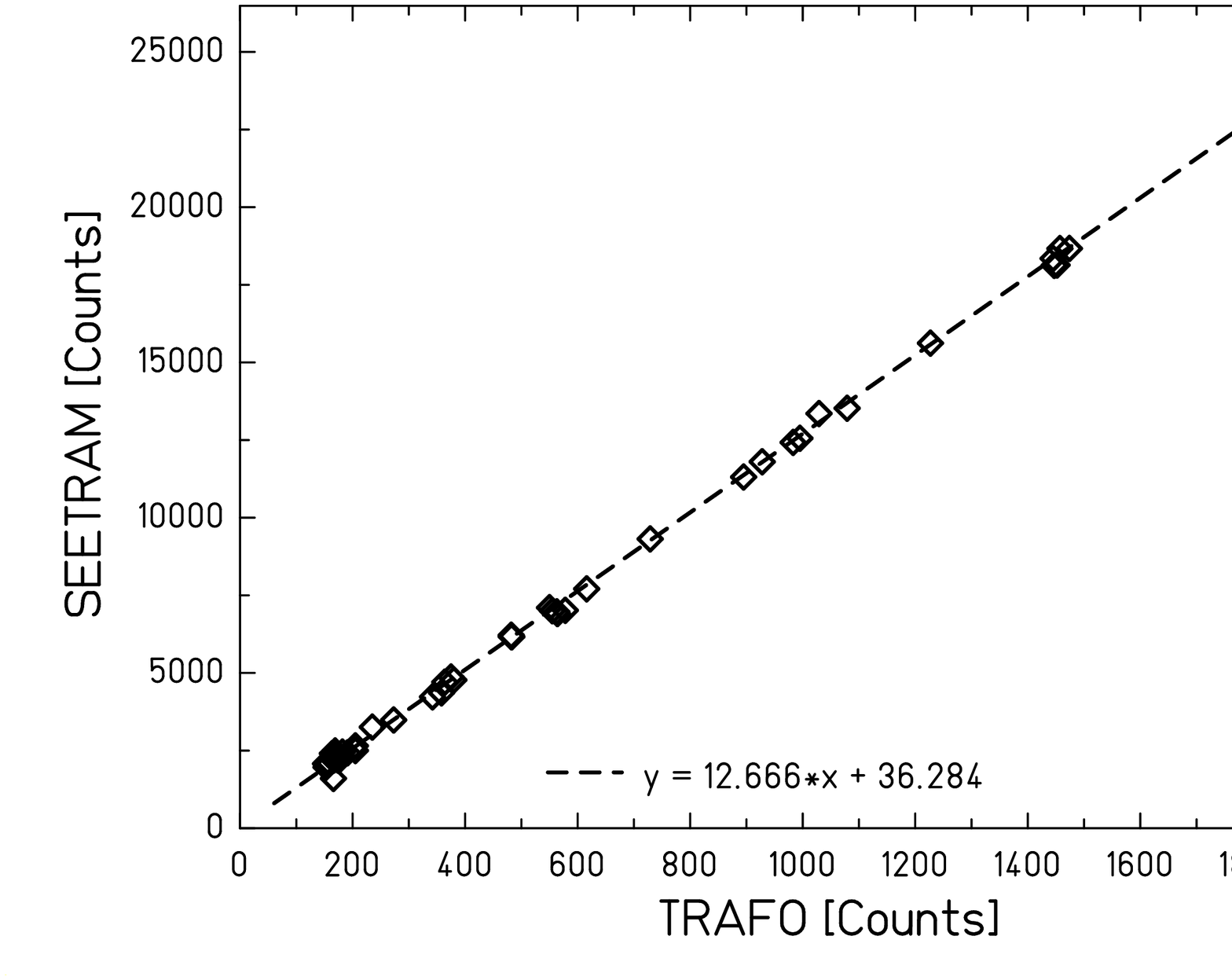}\\
\end{array}$
\end{center}
\caption{\label{fig:trafocalib}Top: Dashed line shows the quadratic calibration curve fitted to the data showing scintillator counts versus the SEETRAM output. Solid line shows the linear part of the fit. Bottom:
Linear calibration fit to the data showing the SEETRAM output
versus the TRAFO output.}
\end{figure}

The normalization to the primary-beam intensity was done by
counting the number of the incoming beam particles using the
signal from the beam-current transformer (TRAFO) \cite{Reeg2001}
used for the SIS beam monitoring. The advantage of using the TRAFO
instead of the standard FRS beam monitor, SEETRAM (Secondary
Electron TRAnsmission Monitor) \cite{Junghans1996312,Jurado2002603}, is the fact that there is no layer of matter introduced to the beam line, which would act as an additional target. However the SEETRAM output served as an intermediate information to connect the absolute calibration with the scintillator to the TRAFO output. SEETRAM measures the electron current as a function of the number of incident beam particles which is measured with a scintillation detector during the calibration run. Due to the saturation of the scintillation detector output at large particle fluxes the calibration was made up to the particle rate in the order of $\sim$10$^5$
particles per second. The calibration data and a quadratic fit are
presented in Fig. \ref{fig:trafocalib}. Since SEETRAM itself
does not suffer any sizable saturation at these particle rates,
the linear term of the quadratic fits where taken as the
calibration factor to compensate the scintillator saturation.
TRAFO could not measure the low intensity particle flux used in the SEETRAM calibration, so TRAFO was calibrated with higher beam intensity against the calibrated SEETRAM output. 
The linear calibration fit of TRAFO is presented in Fig. \ref{fig:trafocalib}.

The next step undertaken before merging the velocities measured in
different magnetic-field settings was the correction for the slight variation of the 
angular transmission as a function of the ion's magnetic rigidities in the first
and second halves of the FRS \cite{Benlliure2002493}. While the heavy residues are produced with rather narrow angular distributions and they are fully transmitted through the FRS, the angular
distributions of light residues are rather broad, and the angular
transmission of these residues may be as low as 10\%. The angular
transmission of the FRS has been under intense investigation in
many experiments in the past. For the case of fragmentation
reactions, a detailed description of the transmission of each ion
species through the magnetic fields of FRS is given in ref.
\cite{Benlliure2002493}. In the same work, also an algorithm for
correcting for the transmission losses due to the limited angular
acceptance is given. This algorithm has been adapted in the present
work. After this correction the velocity distributions closely represent the distributions inside the afore mentioned angular acceptance of the spectrometer (15 mrad around the beam axis). The applied transmission correction factors were assumed to have a relative uncertainty of 15\%. 
In the present experiment, the dead-time of the data-acquisition
system was varying, depending on the counting rate, between 2\%
and 50\%. For each setting, the dead-time values have been
registered, and measured counting rates consequently corrected for.

\begin{figure}
\includegraphics[width=0.5\textwidth]{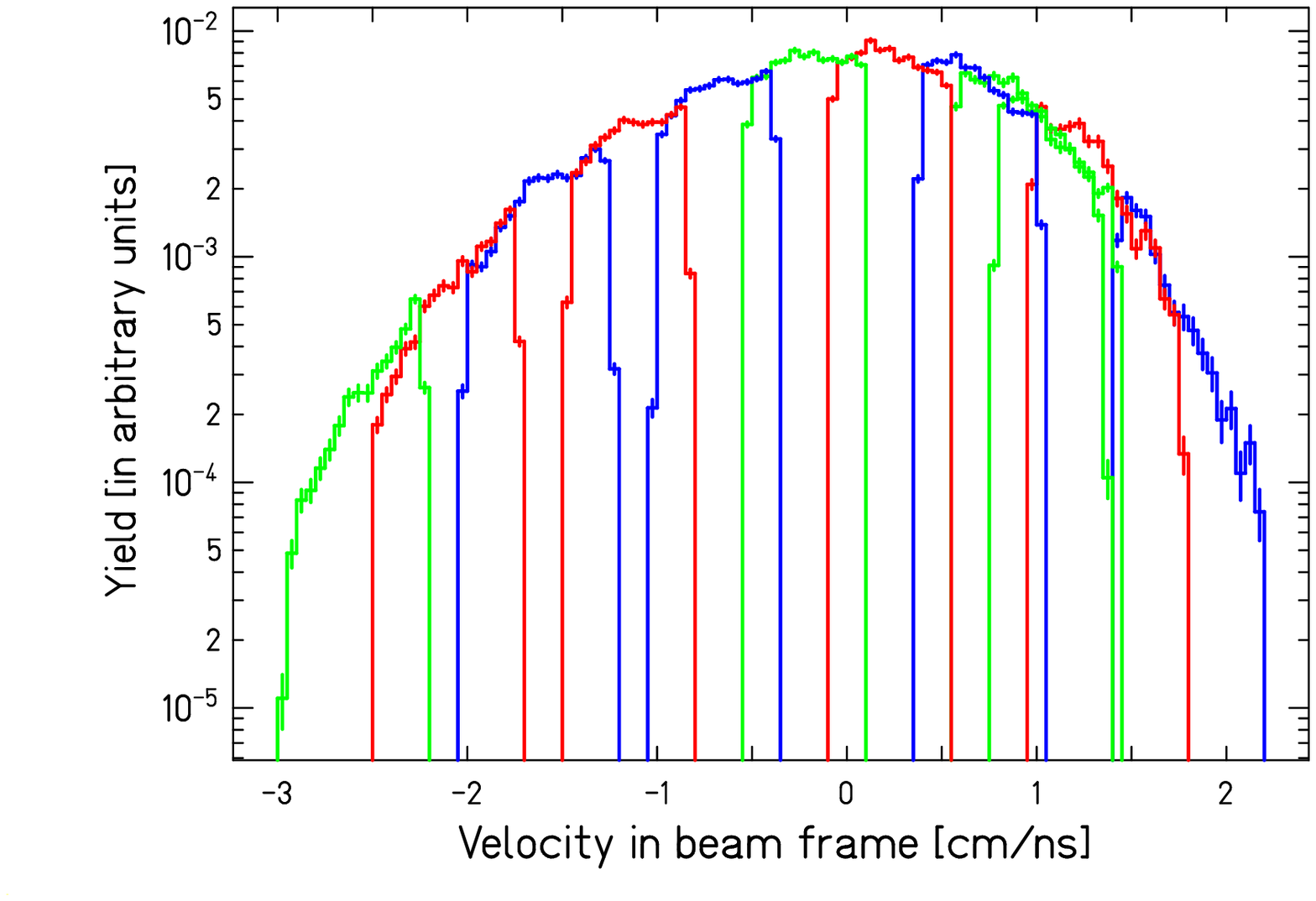}
\caption{\label{fig:ne21sets}(Color online) 
Reconstructed velocity distribution of the nuclide $A$=21, $Z$=10 of the $^{124}$Sn + $^{124}$Sn system inside an acceptance angle of 15 mrad. Contributions of measurements with different magnetic-field settings are shown.}
\end{figure}

Fig. \ref{fig:ne21sets} illustrates the reconstruction of the velocity
distribution of $^{22}$Ne obtained from merging single settings
after performing above-mentioned corrections.

\subsubsection{Determination of the production cross sections}
At this place we will describe how the production cross section of each
nuclide was obtained from its reconstructed velocity distribution.

For the evaluations of the production cross sections, the part of the velocity distributions outside the angular acceptance needed to be estimated. The estimation was based on the isotropy assumption from which it follows that the overall velocity distributions in three-dimensional space have the same standard deviation as the reconstructed longitudinal velocity distributions inside the angular acceptance of the spectrometer. This procedure is especially reliable for the narrow distributions of the heavy residues but the cross sections of the light residues (Z $\lesssim$ 14) may have been somewhat underestimated due to the shape asymmetries in their velocity distributions. 

The integral of the reconstructed velocity distribution of a given
fragment was used as a basis to obtain its production cross
section. This is done in the following way: The yield $Y(A,Z)$ of
a given residue is obtained by integrating its completely
reconstructed velocity distribution. By this method one ensures
that there is no double counting due to the overlap of the
neighboring magnetic-field settings.

The determination of production cross sections $\sigma(A,Z)$ from
the measured yield $Y(A,Z)$ of single nuclide $(A,Z)$ is calculated as:
\begin{equation}
\sigma(A,Z) = \frac{Y(A,Z)\cdot \alpha}{N_{Sn}} ,
\label{eq:crosss}
\end{equation}
where $\alpha$ is a correction factor for the losses due to
secondary reactions in scintillator and degrader, and $N_{Sn}$ is
the number of target nuclei over unit area.

\begin{figure}
\includegraphics[width=0.5\textwidth]{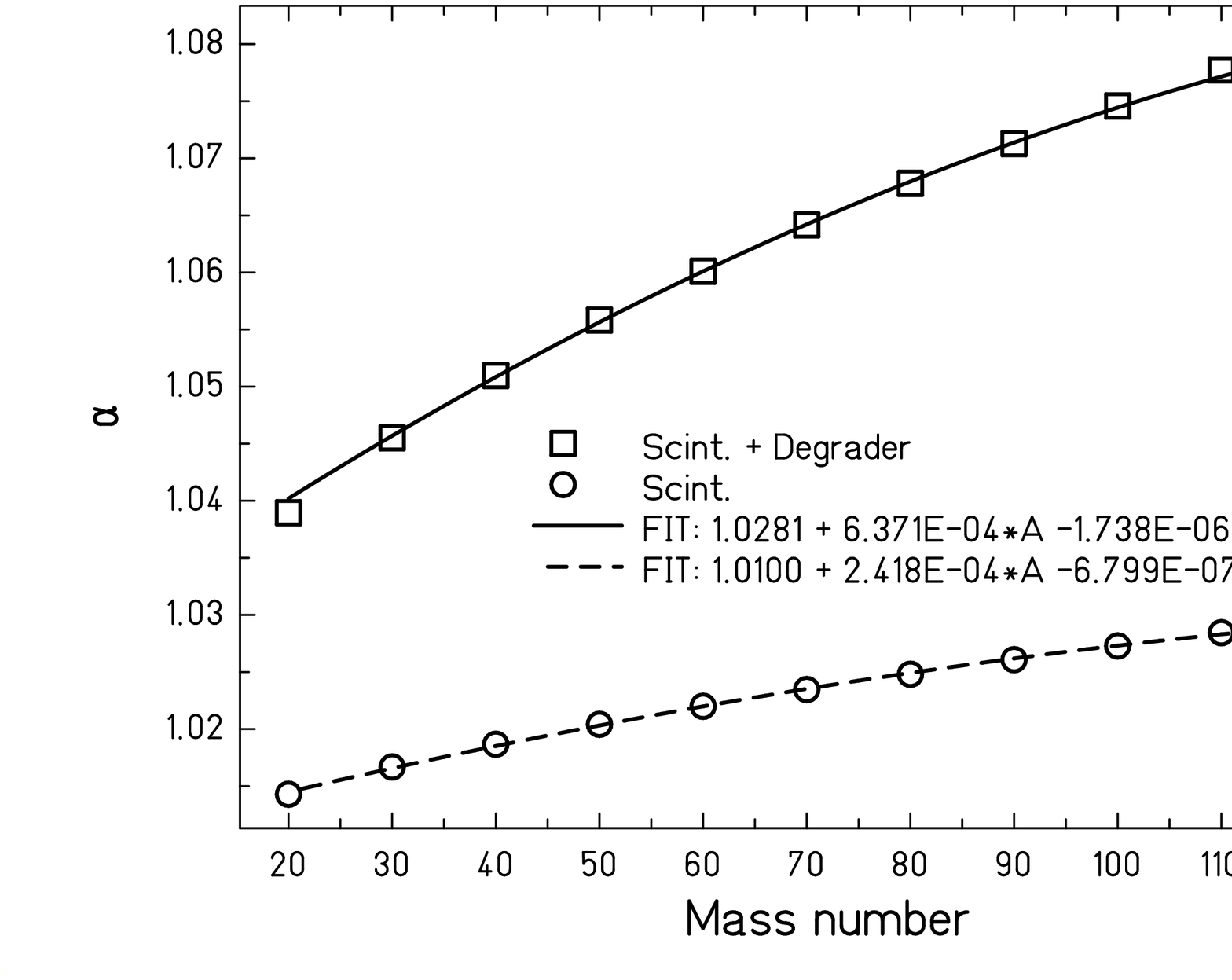}
\caption{\label{fig:secondary}Secondary-reaction correction factor in different layers
of matter obtained from AMADEUS \cite{AMA} calculation.}
\end{figure}

The function $\alpha$, plotted in Fig. \ref{fig:secondary}, is
a quadratic fit made to the correction factors calculated with the 
code AMADEUS \cite{AMA} for every ten atomic mass units. AMADEUS gives the percentage of
nuclear reactions in matter for a given fragment with a given
velocity. Corrections ranging from 0.5\% to 8\% were applied. We
assume a relative uncertainty of 10\% for the correction for
secondary reactions. Please note, that no corrections due to the
secondary reactions in the targets were needed, as both targets
were very thin, and the probability for a fragment to react in one of
the targets was less than 0.1\%.

The number of target nuclei over unit area $N_{Sn}$ is equal to the
number of individual scattering centers per unit volume, $n$,
times the thickness, $x$, of the target:
\begin{equation}
N_{Sn}= n \cdot x = \frac{\rho \cdot N_{Av}}{M} \cdot \frac{T}{\rho}
= \frac{N_{Av}\cdot T}{M}\textrm{,} \label{eq:density}
\end{equation}
where $T$ is the density thickness of the target in mg $\cdot$
cm$^{-2}$, $N_{Av}$ = 6.022$\cdot$10$^{23}$mol$^{-1}$ Avogadro's
number, $M$ the atomic weight of the target material [mg$\cdot$
mol$^{-1}$] and $\rho$ the density [mg$\cdot$ cm$^{-3}$] of the
target material. Numerical values for the quantities in the
equation \ref{eq:density} are given in table \ref{tab:thickvalues} for both tin targets used in the experiment.
\begin{table}[h!]
\centering
\begin{tabular}{|c|ccc|}
    \hline
target &  $T$ [mg$\cdot$ cm$^{-2}$] & $M$ [g$\cdot$ mol$^{-1}$]  & Enrichment [\%]    \\
    \hline
$^{112}$Sn  & 126.7 $\pm$ 0.6 & 112.4 $\pm$ 0.2 & 99.5$\pm$0.2 \\
$^{124}$Sn  & 141.8 $\pm$ 0.7 & 123.9 $\pm$ 0.2 & 97.5$\pm$0.2 \\
    \hline
\end{tabular}
\caption{Numerical values used in the equation \ref{eq:density}.}
\label{tab:thickvalues}
\end{table}

Finally, to ensure that each count in the distribution
corresponding to given $A$ and $Z$ in the identification plot
indeed represents this isotope and does not come from products of
ionic charge-changing processes and secondary reactions in the
detector materials and degrader in the beam line, all spectra were
accumulated under the condition that each measured fragment had
the same ratio $A$/$Q$ in both stages of the FRS. Due to this
constraint on the $A$/$Q$ ratio the ions which undergo ionic
charge-changing processes and which are responsible for the most
of the contaminants can be suppressed efficiently. The amount of
remaining contaminants from background of secondary-reaction
products which were transmitted to the final image plane and
fulfilled the constraint on the $A$/$Q$ and on the energy loss in
the ionization chamber, was estimated to be less then 1\%. More
details can be found in ref. \cite{Enqvist2002435}. 

\section{\label{sec:results}Results and discussion}
With the method described in the previous chapter we obtained the
longitudinal velocity distributions of the projectile residues –
fully identified in mass and atomic number –  in the reactions
$^{112}$Sn + $^{112}$Sn and $^{124}$Sn + $^{124}$Sn at an incident
beam energy 1$\cdot A$ GeV. The width and the mean value of the
velocity distributions as well as the production cross sections
were determined, and will be presented.

\subsection{Velocity distributions and their moments}
\begin{figure}
\includegraphics[width=0.5\textwidth]{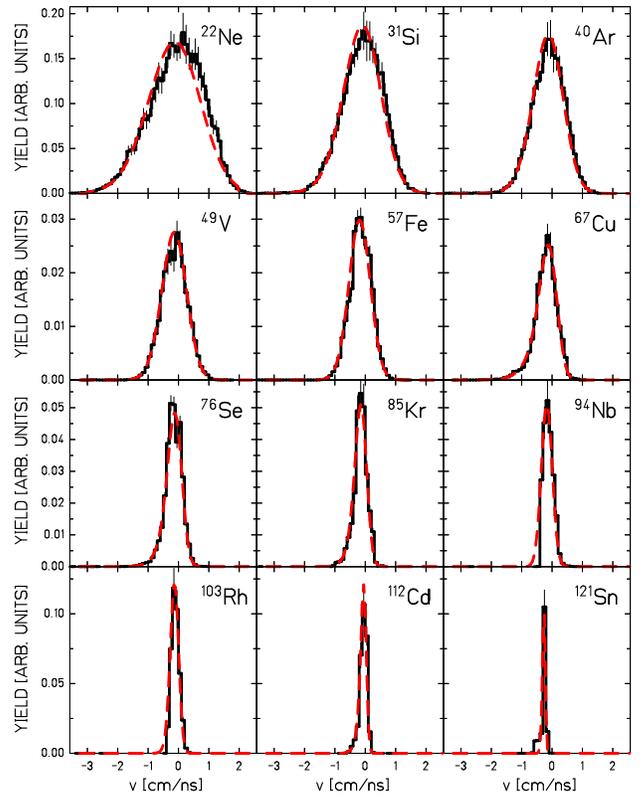}
\caption{\label{fig:124ex}(Color online) Velocity
distributions in the rest frame of the projectile for several
nuclides measured in the reaction $^{124}$Sn + $^{124}$Sn. The fitted functions are shown as dashed red lines. Distributions represent velocities inside an angular range of 15 mrad.}
\end{figure}

\begin{figure}
\includegraphics[width=0.5\textwidth]{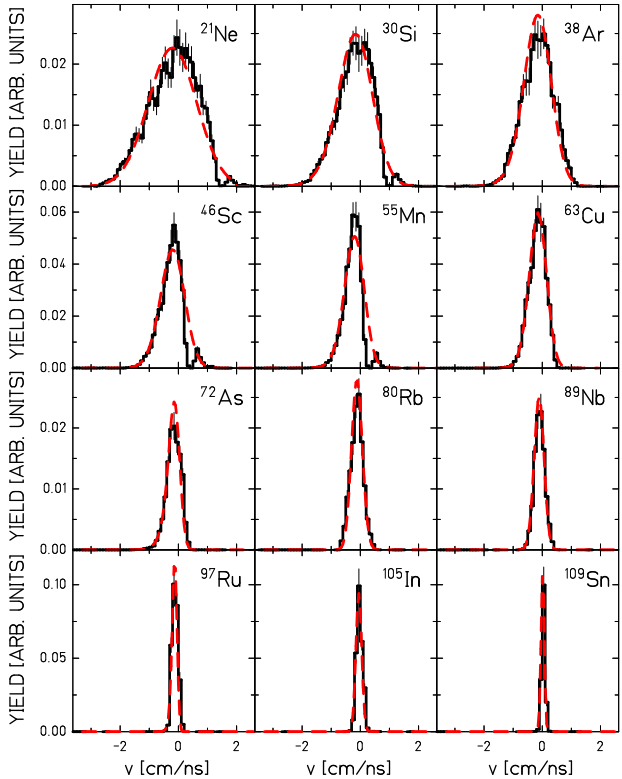}
\caption{\label{fig:112ex}(Color online) Velocity
distributions in the rest frame of the projectile for several
nuclides measured in the reaction $^{112}$Sn + $^{112}$Sn. The fitted functions are shown as dashed red lines. Distributions represent velocities inside an angular range of 15 mrad.}
\end{figure}

In Fig.~\ref{fig:124ex} and in Fig.~\ref{fig:112ex} some examples of the measured distributions of longitudinal velocity inside an angular acceptance range of 15 mrad around the beam
direction in the frame of the projectile are presented for few selected nuclides.
By observing the general shapes of the velocity distributions and
without going into details of different reaction mechanisms, one
can explain the general features in terms of abrasion-ablation
model \cite{Gaimard1991709}. In this model the heavy fragments shown in Fig.~\ref{fig:124ex}
and in Fig.~\ref{fig:112ex} are produced in
peripheral collisions with small overlap between the projectile
and the target nucleus. This results in a formation of slightly
excited prefragments, which then de-excite via
evaporation of neutrons, light charged particles and light
clusters. As the number of abraded nucleons is small, the fluctuations in the velocity distribution of created prefragments are also small \cite{Goldhaber1974306}, and consequent evaporation stage only slightly increases the width of the velocity distribution. 
Thus, these residues show narrow velocity distributions with the
mean value only slightly lower than that of the beam particles.
Lighter residues are presumably produced at smaller impact
parameters, where the introduced excitation energy can be high enough for
thermal instabilities to set in \cite{Botvina1995737,Schuttauf1996457}.
These residues portray wider velocity distributions, indicating a
larger number of abraded and evaporated nucleons. For all the
residues, the longitudinal velocity distributions show
Gaussian-like shape with a slightly enhanced tail in the slower
side in case of the lighter residues. This asymmetry in the shape of
velocity distributions of the lightest residues shows, as discussed
in references \cite{Ricciardi:2005ht,PhysRevC.70.054607,PhysRevC.78.044616} that these fragments have been produced via different reaction mechanisms like e.g. simultaneous and/or sequential decay.

For the sake of more quantitative analysis, the distributions were
fitted with one Gaussian with an exponential tail. This fitting
function was chosen because it resembled the experimental data
sufficiently well. Advantage of the fitting was to get rid of some
unwanted features of the distributions that were introduced only
due to experimental limitations. Some of the velocity
distributions show, especially in the case of $^{112}$Sn +
$^{112}$Sn, cuts caused by slits, which were inserted to protect
the detectors from the primary beam and its first two charge
states; see e.g. $^{21}$Ne, $^{30}$Si, $^{46}$Sc or
$^{55}$Mn in Fig.~\ref{fig:112ex}. The fitting procedure could, thus,
recover some of the incompletely measured velocity distributions. The uncertainty introduced by this fitting procedure is typically of the order of 2\% for each extracted parameter.

\begin{figure}
\begin{center}
$\begin{array}{c}
\includegraphics[width=0.45\textwidth]{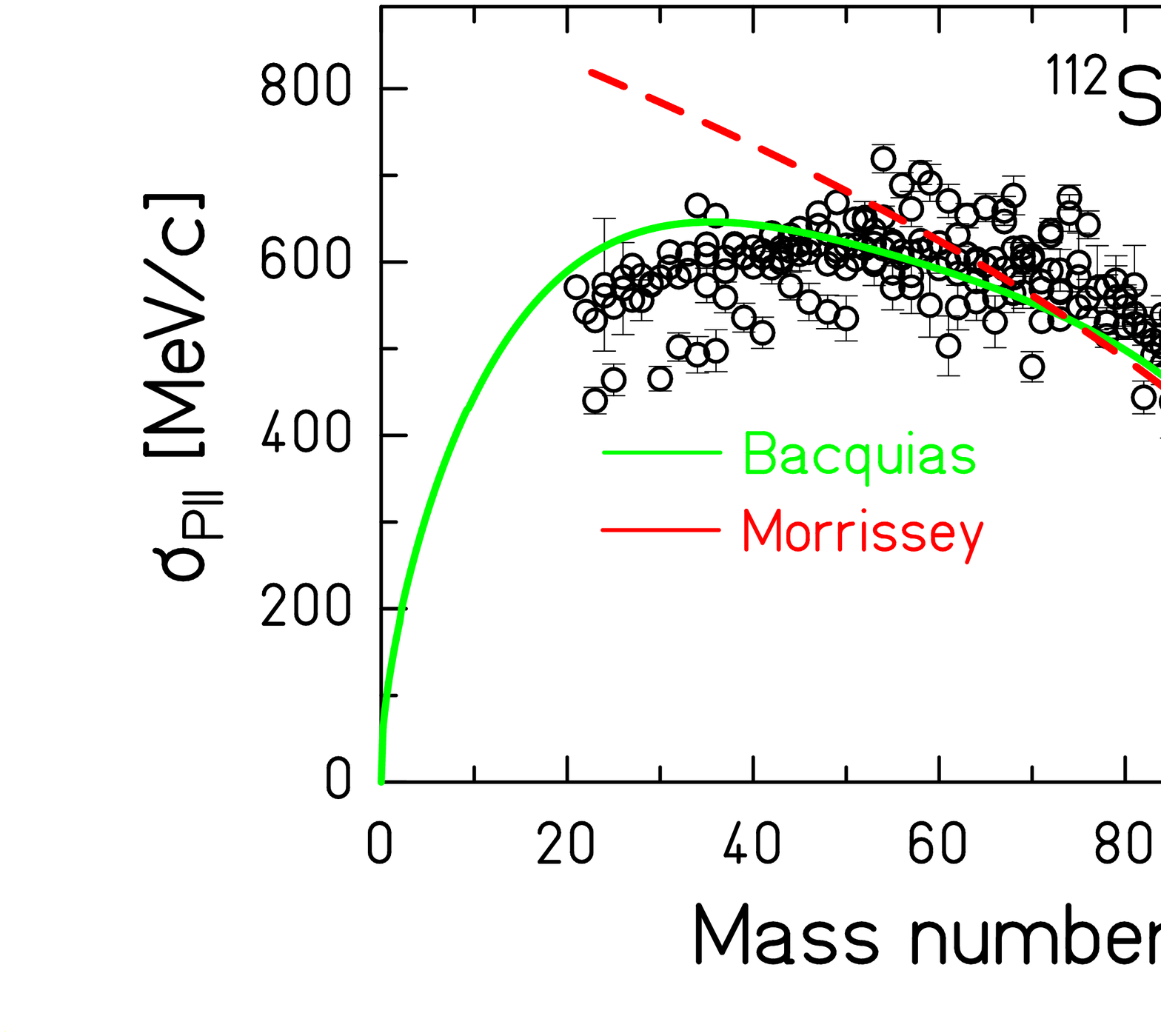}\\
\includegraphics[width=0.45\textwidth]{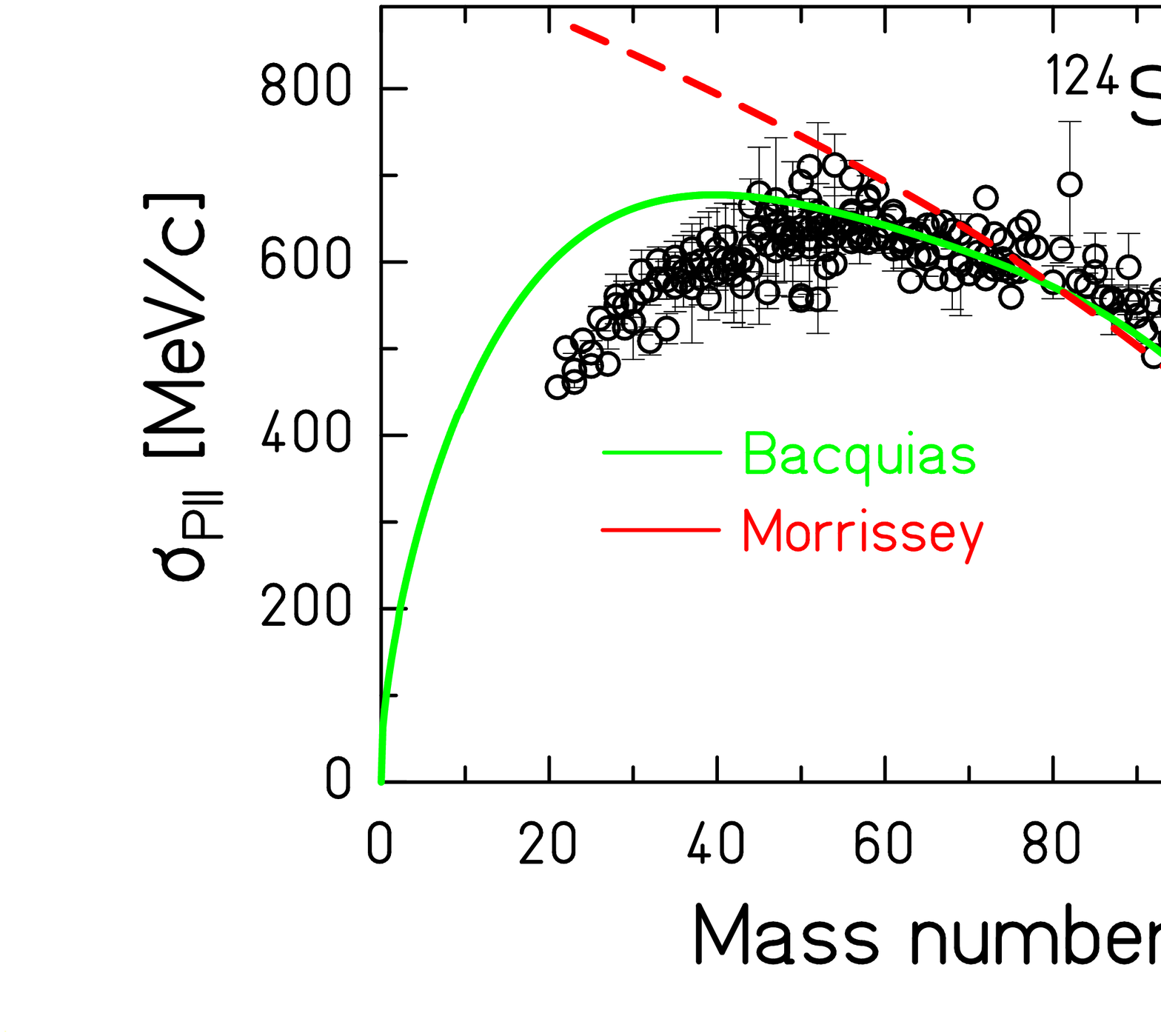}\\ 
\end{array}$
\end{center}
\caption{\label{fig:widths}Standard deviation of the Gaussian part of the fitted functions to the
fragment momentum distributions for both systems;
$^{112}$Sn(1$\cdot A$ GeV)+$^{112}$Sn (upper diagram) and
$^{124}$Sn(1$\cdot A$ GeV)+$^{124}$Sn (lower diagram). The values
are given in momentum units [MeV/c]. Given curves present the
prediction of theoretical model \cite{Bacquias} and the
empirical parametrization by Morrissey \cite{PhysRevC.39.460}.}
\end{figure}

In Fig.~\ref{fig:widths} the width, standard deviation
$\sigma_\|$ of the fitted Gaussian function with exponential tail, is
given for the longitudinal momentum distributions for all
fragments measured in two reactions. Also shown are a theoretical prediction
\cite{Bacquias} and the empirical parametrization of Morrissey \cite{PhysRevC.39.460}.
From these figures we see that the width of the measured longitudinal momentum distributions
first increase with decreasing mass of the final residue. The maximum is reached for the final-fragment mass close to half the mass of the projectile. For lower masses, the width then decreases.
Uncertainties for the standard deviations are given by the fitting routine which calculates
them according to the uncertainties of each bin content of the velocity distributions. The total uncertainty for each bin content were calculated from individual error sources by using the error propagation law.  This uncertainty consist of both statistical and systematic. Other uncertainties that might have not been considered have no sizable contribution, when added quadratically to the estimated ones, since they are in total less than 3\%.

According to the statistical model of Goldhaber \cite{Goldhaber1974306}, the longitudinal
momentum of the projectile-like fragments after the first reaction
stage are determined by the intrinsic Fermi motion of the
constituent nucleons which are removed from the projectile during the
abrasion process. In this model the individual nucleons within the
projectile have their own momenta that sum up to zero in the rest
frame of the projectile. The abrasion stage then removes nucleons
with no preference with respect to their momentum, and the sum of the momenta of
the remaining nucleons in the prefragment may not sum up to zero
anymore. Due to the momentum conservation, the sum of momenta of abraded
nucleons, has to be opposite to the momentum of the prefragment. More
nucleons are removed in abrasion more fluctuations of the remaining
total momentum may occur, and the maximum is reached for masses equal the half of the mass 
of the projectile. By this model, the projectile remnants end up with three-dimensional Gaussian-shape momentum distributions, whose widths are determined by the number
of removed nucleons. Although it considers only the abrasion stage, the Goldhaber's model  \cite{Goldhaber1974306} has often been used in interpreting experimental results.
Recently, Goldhaber's model has been refined and the influence of different decay stages, i.e. simultaneous and/or sequential decay, has been incorporated in the description of
the momentum dispersion of final fragmentation residues \cite{Bacquias}. This has resulted in an improved description of the width of the momentum distribution of fragmentation residues.

The projectile remnant subsequently enters, depending on the excitation energy induced 
in the abrasion stage, the stage of simultaneous and/or sequential decay. These decay stages introduce
additional fluctuations in the momentum distribution as discussed in ref. \cite{Bacquias}.

Another often used description of the momentum width is based on the empirical approach of Morrissey \cite{PhysRevC.39.460}. Although this approach describes very well the width of the momentum distribution close to the projectile, it fails severely for the masses smaller than the half of the projectile mass, see Fig.~\ref{fig:widths}.

\begin{figure}
\includegraphics[width=0.5\textwidth]{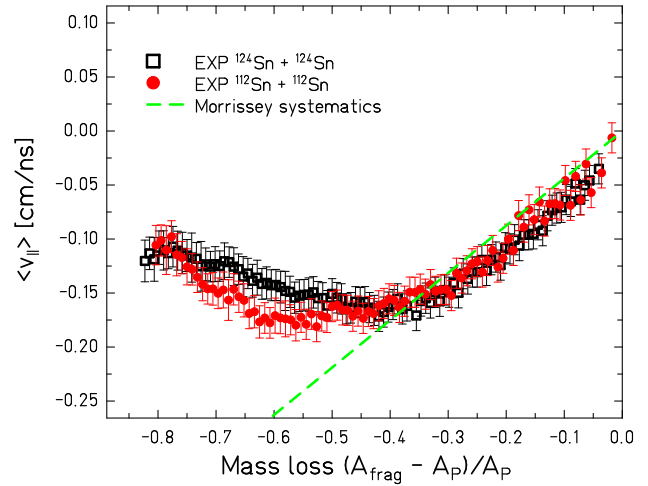}
\caption{\label{fig:meanv}(Color online) Mean value
of the longitudinal velocity distributions in the frame of the
projectile of residues produced in the reactions $^{124}$Sn
(1$\cdot A$ GeV)+ $^{124}$Sn (open squares) and $^{112}$Sn (1$\cdot A$ GeV) + $^{112}$Sn  (filled circles) as a function of their relative mass
loss. The mean values represent the mean velocities inside an angular range of 15 mrad.}
\end{figure}

Another characteristic of the velocity distribution is its mean
value. In Fig.~\ref{fig:meanv} we present the average mean velocity in
the frame of the projectile for each mass number of the fragmentation residues ($A_{frag}$)
produced in the reactions induced by $^{124}$Sn and $^{112}$Sn, as
a function of their relative mass loss defined as
$(A_{frag}-A_{proj})/A_{proj}$. Uncertainties of the mean values are also based on the uncertainties of each bin of the velocity distributions and were given by the fitting routine. In addition to this, the uncertainty of the mean value contains the uncertainty in determining the velocity of the primary beam since the mean values are given in the beam frame. 

In Fig.~\ref{fig:meanv} the dashed line shows the expected mean velocities according to the systematics of Morrissey~\cite{PhysRevC.39.460}.
The mean value represents the overall velocity shift induced in
these reactions. A similar pattern is observed in both systems.
Residues close to the projectile show a clear decrease of the mean
velocity with their mass loss which closely follows the systematics of
Morrissey~\cite{PhysRevC.39.460}. For this region, there is also no difference between the mean velocities in the two systems. The Morrissey systematics does
not contain any interpretation about the reaction mechanism, it is
just a fit made to the available data at that time. Nevertheless, this behavior can be explained by simple two-body interaction, namely friction, between the projectile and target nuclei in peripheral heavy-ion collisions. Friction appears as a consequence of
interactions between the projectile and target matter in the overlapping region, and leads to a
conversion of relative kinetic energy into excitation energy of projectile and
target spectators. Due to this loss in kinetic energy, the velocity of spectator
residues is slightly shifted towards the velocity of the reaction partner,
i.e. projectile residues are slowed down \cite{AbulMagd1976327,Weber1994659}.

However, in less peripheral collisions, the two-body kinematics is no longer applicable since there occurs a formation of participant zone and the projectile and target spectators emit particles. 
In Fig.~\ref{fig:meanv} at relative mass losses around 0.4, corresponding to $A
\approx$ 67 and $A \approx$ 74 in  $^{112}$Sn + $^{112}$Sn and in
$^{124}$Sn + $^{124}$Sn, respectively, the mean velocity levels off
and for large mass losses the mean velocity starts to rise again. 
This is in clear contrast with the friction picture, according to which one would expect more kinetic-energy dissipation as the mass of the residue is decreasing and, thus, lower velocity. As discussed in refs. \cite{PhysRevC.64.034601,PhysRevLett.90.212302} this
leveling-off and increase in the mean velocity of the final
residue with decreasing mass can be the evidence of the influence
of the participant blast on the properties of the projectile-like
spectator. For the lowest masses with more than 0.5 relative mass loss the data suggest that there is also a small deviation between the two systems.  

At this place, we would like to make a comment concerning the influence of limited angular acceptance of the FRS. As we have said above, the angular transmission varies as a function of the ion's magnetic rigidities in the first and second halves of the FRS. After correcting for this effect, the velocity distributions (see Figs.~\ref{fig:124ex} and \ref{fig:112ex}) as well as mean values (see Fig.~\ref{fig:meanv}) represent fragment velocity inside an angular range of 15 mrad. This is very important to keep in mind, when comparing our results with experiments performed with full-acceptance set-up. In our case, only those fragments which are emitted with rather small angles, i.e. 15 mrad around the beam axis, are measured and thus the measured velocity distributions presented here are only slightly influenced by e.g. binary-type events in which due to a strong Coulomb repulsion the produced fragments are emitted with larger angles \cite{PhysRevC.70.054607,PhysRevC.76.051603,Ricciardi:2005ht}. On the contrary, in the full acceptance experiments, all products, regardless of their production mechanism, are detected and this of course lead to somewhat different shape of the velocity distribution as well as to lower average velocities. Due to this fact, effects seen in e.g. Fig.~\ref{fig:meanv} are not easy to be observed in full-acceptance experiments. Discussion of this effect and comparison with
theoretical predictions is a topic of a forthcoming publication and beyond the scope of the present paper.

\subsection{Production cross sections}
\begin{figure}
\begin{center}
$\begin{array}{c}
\includegraphics[width=0.45\textwidth]{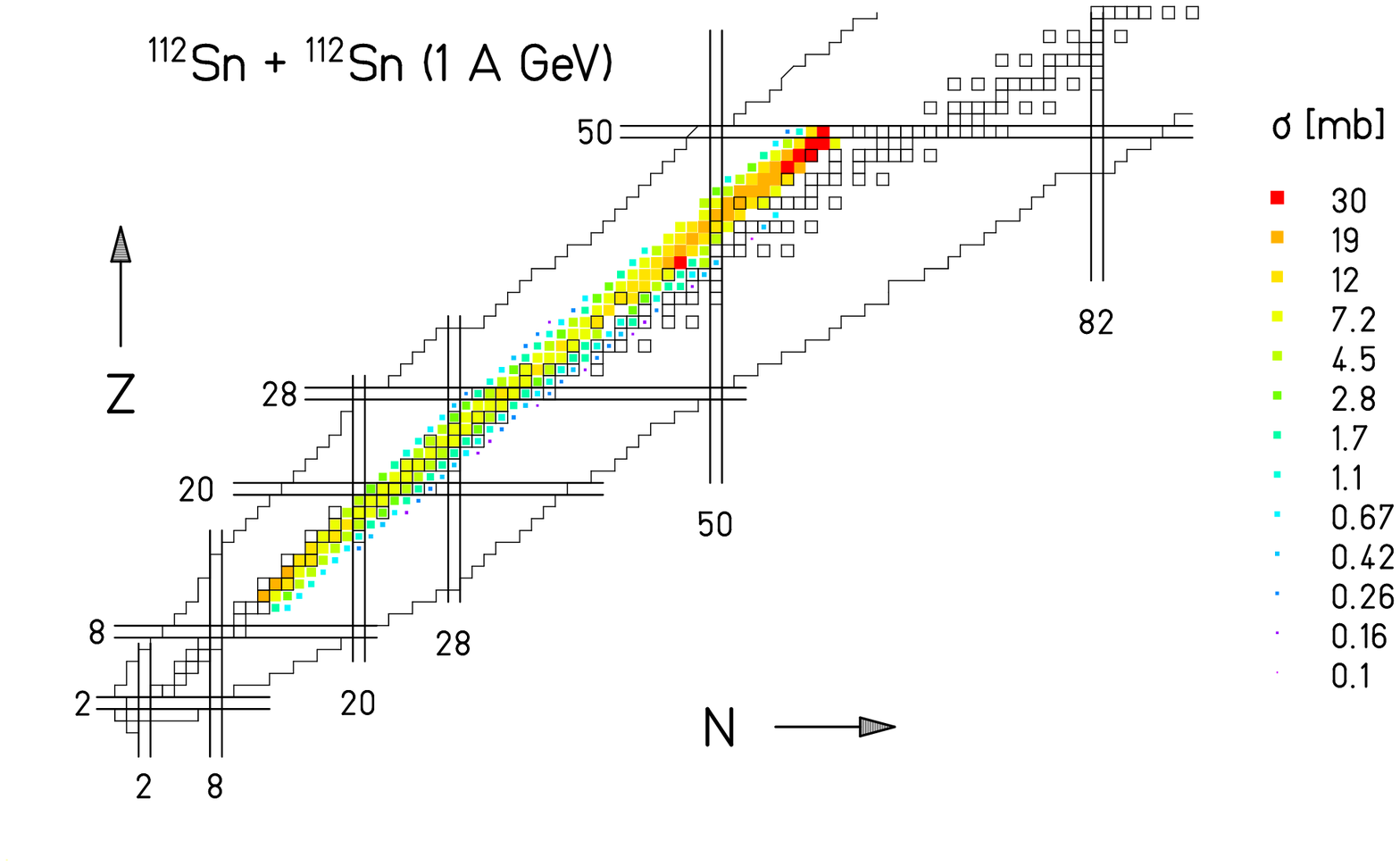}\\
\includegraphics[width=0.45\textwidth]{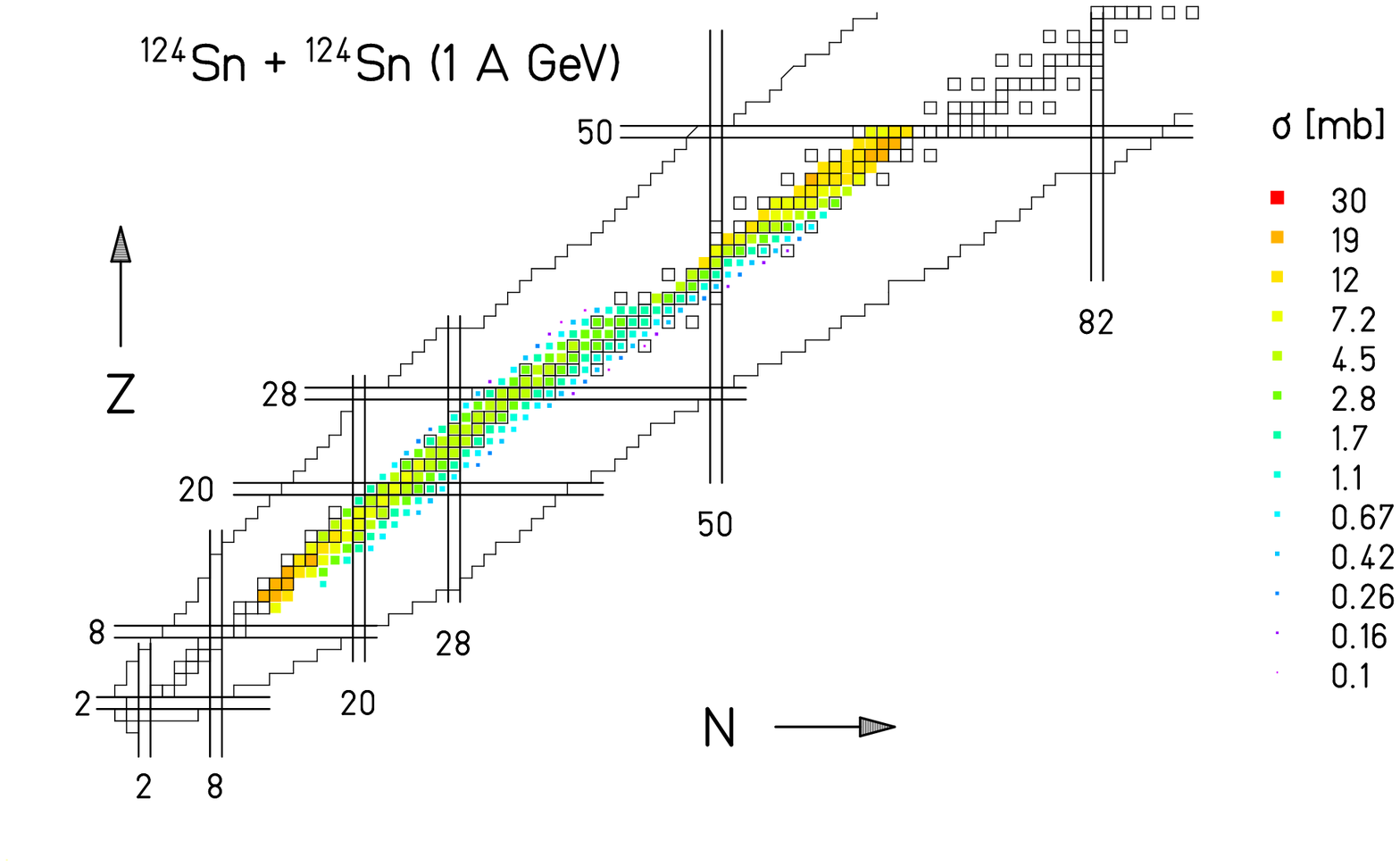}\\ 
\end{array}$
\end{center}
\caption{\label{fig:crossecs}(Color online) Cross sections for both systems; $^{112}$Sn(1$\cdot A$ GeV)+$^{112}$Sn (upper diagram) and $^{124}$Sn(1$\cdot A$ GeV)+$^{124}$Sn (lower diagram).} 
\end{figure}

As already discussed, the velocity distributions served also to
determine the production cross sections of the measured fragments.
An overview of the measured production cross sections presented on
the chart of nuclides is shown in Fig. \ref{fig:crossecs}. For
several isotopes the cross sections could not be determined due to
lack of statistics or because of severe cuts in the velocity
distributions, or simply because of the limited range of the
magnetic rigidity that was measured. Measured isotopic distributions are also
plotted in Fig.~\ref{fig:crossXwepax} and numerical values of the production cross sections are given in the Appendix. Uncertainties of the cross sections are based on the uncertainty of the integral of the velocity distributions given by the fitting routine. An additional uncertainty was introduced by the estimation of the parts of the velocity distributions outside the angular acceptance. Fig.~\ref{fig:crossXwepax} also shows the cross sections obtained from the EPAX parametrization \cite{PhysRevC.61.034607}. EPAX is a semi-empirical parametrization of the cross sections of heavy residues from fragmentation reactions based on the idea that fragmentation products result from long, sequential evaporation chains, at the end of which the evaporation attractor line is reached. Fragment cross sections obtained in both experiments agree with the EPAX parametrization reasonably well.
\begin{figure}
\begin{center}
\includegraphics[width=0.5\textwidth]{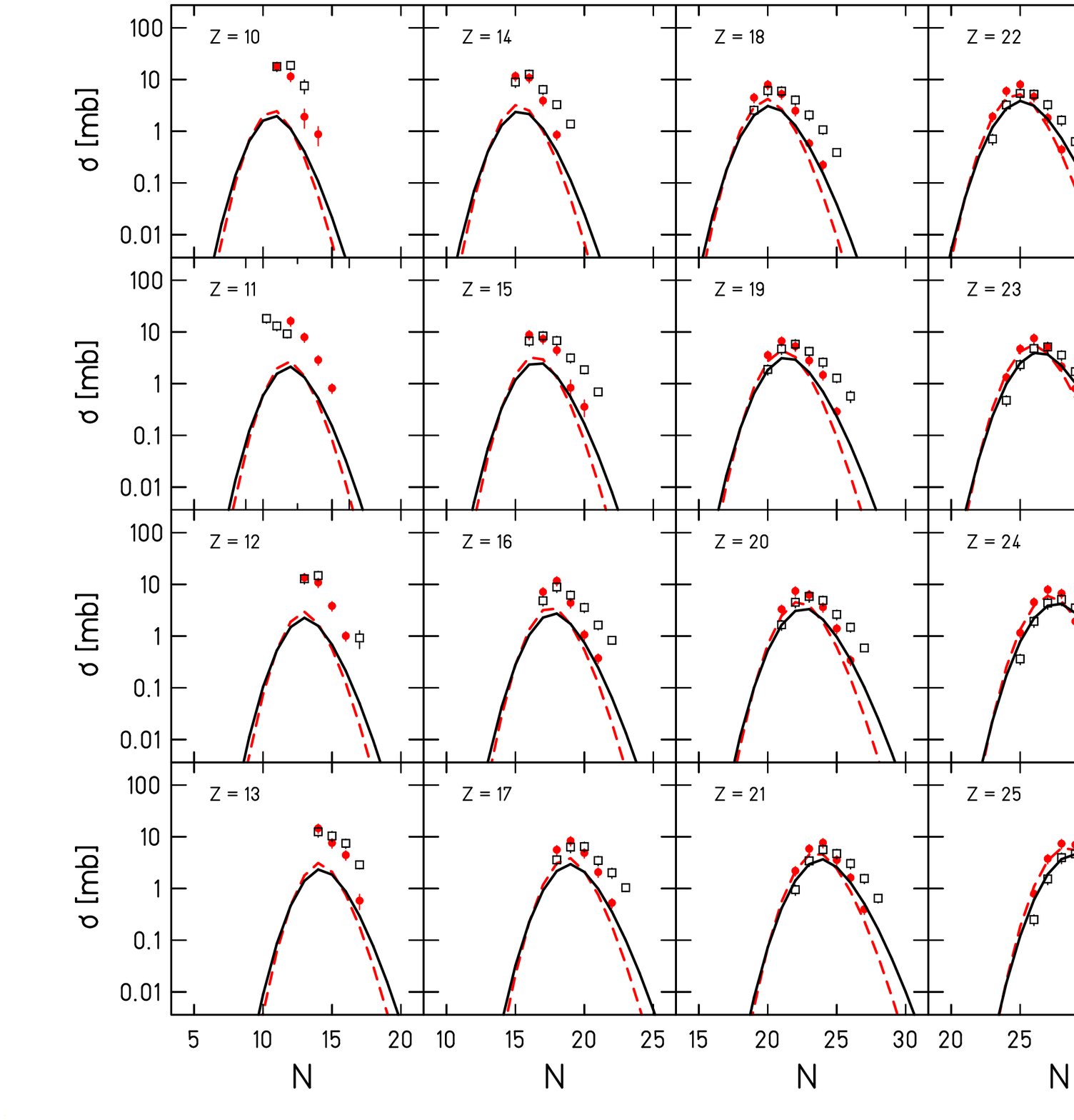}
\end{center}
\end{figure}
\begin{figure}
\begin{center}
\includegraphics[width=0.5\textwidth]{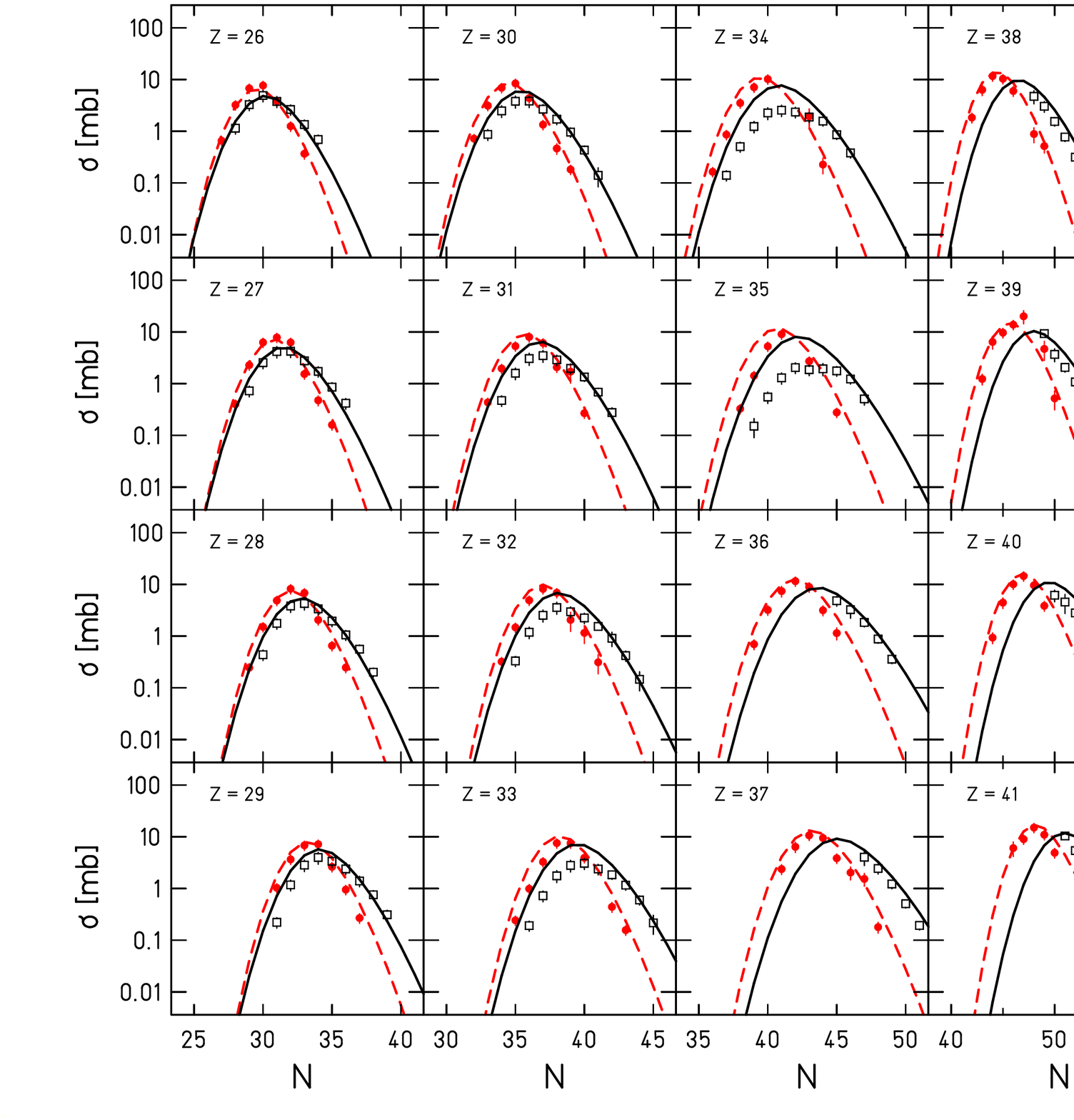}
\end{center}
\end{figure}
\begin{figure}[ht]
\begin{center}
\includegraphics[width=0.5\textwidth]{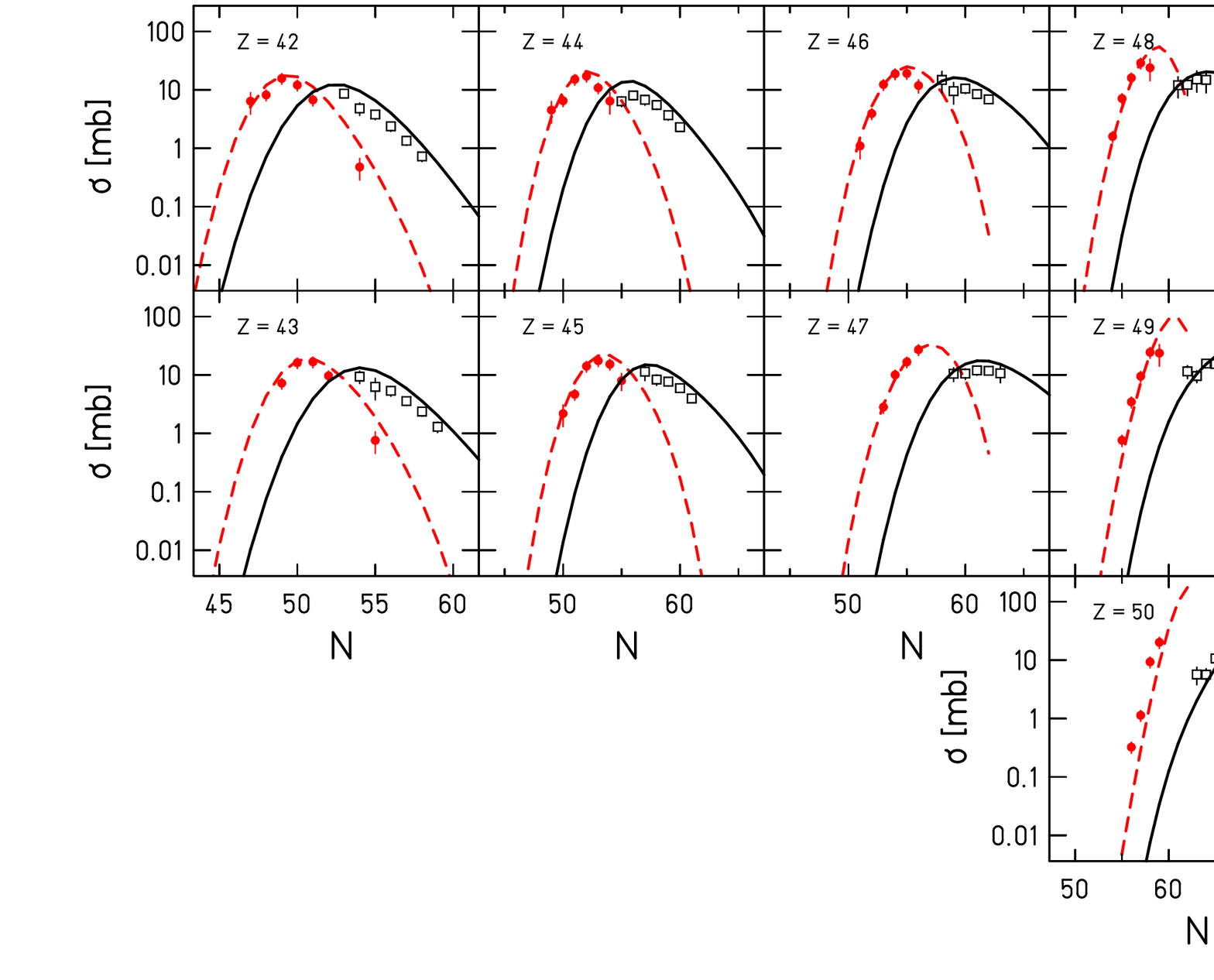}
\end{center}
\caption{\label{fig:crossXwepax}(Color online) Isotopic cross sections of the measured fragments in reaction $^{112}$Sn + $^{112}$Sn at 1$\cdot A$ GeV (filled dots) and in $^{124}$Sn + $^{124}$Sn at 1$\cdot A$ GeV (open squares). Dashed lines represents the prediction of EPAX \cite{PhysRevC.61.034607} for $^{112}$Sn + $^{112}$Sn and solid lines for $^{124}$Sn + $^{124}$Sn.} 
\end{figure}

\section{\label{sec:conclusions}Conclusions}
The high-resolving-power magnetic spectrometer, FRS, was used to
measure  the longitudinal velocity of the residues  produced in
the reactions $^{124}$Sn + $^{124}$Sn and $^{112}$Sn + $^{112}$Sn
at 1$\cdot A$ GeV with a relative uncertainty of $\sim 1 \times
10^{-4}$. This precision allows to investigate the mechanisms
responsible for fragment formations. The mean value of the
longitudinal velocity distributions of light projectile-like
residues show a clear deviation from what one would expect on the
basis of the friction picture in heavy-ion collisions.

The width of the longitudinal velocity distributions shows to deviate from the earlier empirical prediction by Morrissey~\cite{PhysRevC.39.460} and is better reproduced for broader range of data by the modified Goldhaber model \cite{Bacquias} which includes additional corrections to the momentum distributions due to different stages of decay.

The production cross sections were determined from the reconstructed velocity
distributions. The cross sections range over several orders of
magnitude from $\sim$100 $\mu$b to $\sim$30 mb with a relative uncertainty
corresponding to around 20\% in most cases.

\acknowledgments

This work was supported in part by the The Magnus Ehrnrooth foundation.

\appendix

\section{\label{appA} Measured data in reaction $^{124}$Sn + $^{124}$Sn and $^{112}$Sn + $^{112}$Sn at 1$\cdot A$ GeV.}
\begin{longtable*}{ll@{\hspace{5.5mm}}l@{\hspace{6.5mm}}ll@{\hspace{5.5mm}}l@{\hspace{6.5mm}}ll@{\hspace{5.5mm}}l@{\hspace{6.5mm}}ll@{\hspace{5.5mm}}l}
\caption[Production cross sections]{Production cross section of nuclides produced in reactions $^{112}$Sn + $^{112}$Sn at 1$\cdot A$ GeV. Data refers to the full production, corrected for the
limited angular acceptance of the FRS.} \label{112xstable} \\

\hline \hline \\[-2ex]
  \textbf{$Z$} & 
  \textbf{$A$} & 
  \textbf{$\sigma$} [mb]& 
  \textbf{$Z$} & 
  \textbf{$A$} & 
  \textbf{$\sigma$} [mb]&
  \textbf{$Z$} & 
  \textbf{$A$} & 
  \textbf{$\sigma$} [mb]&
  \textbf{$Z$} & 
  \textbf{$A$} & 
  \textbf{$\sigma$} [mb]\\[0.5ex] \hline
   \\[-1.8ex]
\endfirsthead

\multicolumn{12}{c}{{\tablename} \thetable{} -- Continued} \\[0.5ex]
  \hline \hline \\[-2ex]
  \textbf{$Z$} & 
  \textbf{$A$} & 
  \textbf{$\sigma$} [mb]& 
  \textbf{$Z$} & 
  \textbf{$A$} & 
  \textbf{$\sigma$} [mb]&
  \textbf{$Z$} & 
  \textbf{$A$} & 
  \textbf{$\sigma$} [mb]&
  \textbf{$Z$} & 
  \textbf{$A$} & 
  \textbf{$\sigma$} [mb]\\[0.5ex] \hline
  \\[-1.8ex]
\endhead


  \\[-1.8ex] \hline \hline
\endlastfoot
10   &  21  &    18  $\pm$     4  &   22   &  49  &   1.8  $\pm$   0.4  &  31   &  66  &     5  $\pm$     1 &  39   &  89  &   0.5  $\pm$   0.2     \\
10   &  22  &    11  $\pm$     2  &   22   &  50  &  0.44  $\pm$  0.09  &  31   &  67  &     8  $\pm$     2 &  40   &  84  &   0.9  $\pm$   0.2     \\
10   &  23  &   1.9  $\pm$   0.8  &   23   &  47  &   1.3  $\pm$   0.3  &  31   &  68  &     6  $\pm$     1 &  40   &  85  &     4  $\pm$     1     \\
10   &  24  &   0.9  $\pm$   0.4  &   23   &  48  &   4.7  $\pm$   0.9  &  31   &  69  &   2.1  $\pm$   0.4 &  40   &  86  &    10  $\pm$     2     \\
11   &  23  &    16  $\pm$     3  &   23   &  49  &     8  $\pm$     2  &  31   &  70  &   1.7  $\pm$   0.7 &  40   &  87  &    14  $\pm$     3     \\
11   &  24  &     8  $\pm$     2  &   23   &  50  &     5  $\pm$     1  &  31   &  71  &  0.27  $\pm$  0.05 &  40   &  88  &    10  $\pm$     2     \\
11   &  25  &   2.9  $\pm$   0.6  &   23   &  51  &   2.2  $\pm$   0.4  &  32   &  66  &  0.32  $\pm$  0.06 &  40   &  89  &   3.8  $\pm$   0.8     \\
11   &  26  &   0.8  $\pm$   0.2  &   23   &  52  &   0.8  $\pm$   0.2  &  32   &  67  &   1.5  $\pm$   0.3 &  41   &  87  &     6  $\pm$     2     \\
12   &  25  &    13  $\pm$     3  &   23   &  53  &  0.19  $\pm$  0.04  &  32   &  68  &   4.9  $\pm$   1.0 &  41   &  88  &     9  $\pm$     2     \\
12   &  26  &    11  $\pm$     2  &   24   &  49  &   1.2  $\pm$   0.2  &  32   &  69  &     8  $\pm$     2 &  41   &  89  &    15  $\pm$     3     \\
12   &  27  &   3.8  $\pm$   0.8  &   24   &  50  &   4.5  $\pm$   0.9  &  32   &  70  &     7  $\pm$     1 &  41   &  90  &    11  $\pm$     2     \\
12   &  28  &   1.0  $\pm$   0.2  &   24   &  51  &     8  $\pm$     2  &  32   &  71  &   2.1  $\pm$   0.8 &  41   &  91  &     5  $\pm$     1     \\
13   &  27  &    15  $\pm$     3  &   24   &  52  &     7  $\pm$     1  &  32   &  72  &   1.2  $\pm$   0.4 &  41   &  94  &  0.15  $\pm$  0.04     \\
13   &  28  &     8  $\pm$     2  &   24   &  53  &   1.9  $\pm$   0.4  &  32   &  73  &   0.3  $\pm$   0.1 &  42   &  89  &     6  $\pm$     3     \\
13   &  29  &   4.4  $\pm$   0.9  &   24   &  54  &   1.0  $\pm$   0.2  &  33   &  68  &  0.24  $\pm$  0.05 &  42   &  90  &     8  $\pm$     2     \\
13   &  30  &   0.6  $\pm$   0.2  &   24   &  55  &  0.22  $\pm$  0.04  &  33   &  69  &   1.0  $\pm$   0.2 &  42   &  91  &    16  $\pm$     3     \\
14   &  29  &    12  $\pm$     2  &   25   &  51  &   0.8  $\pm$   0.2  &  33   &  70  &   3.3  $\pm$   0.7 &  42   &  92  &    12  $\pm$     2     \\
14   &  30  &    11  $\pm$     2  &   25   &  52  &   3.8  $\pm$   0.8  &  33   &  71  &     8  $\pm$     2 &  42   &  93  &     7  $\pm$     1     \\
14   &  31  &   3.9  $\pm$   0.8  &   25   &  53  &     7  $\pm$     1  &  33   &  72  &     8  $\pm$     2 &  42   &  96  &   0.5  $\pm$   0.2     \\
14   &  32  &   0.9  $\pm$   0.2  &   25   &  54  &     7  $\pm$     1  &  33   &  73  &   3.9  $\pm$   0.8 &  43   &  92  &     7  $\pm$     1     \\
15   &  31  &     9  $\pm$     2  &   25   &  55  &   2.8  $\pm$   0.6  &  33   &  75  &  0.44  $\pm$  0.09 &  43   &  93  &    16  $\pm$     3     \\
15   &  32  &     7  $\pm$     1  &   25   &  56  &   1.1  $\pm$   0.2  &  33   &  76  &  0.16  $\pm$  0.03 &  43   &  94  &    17  $\pm$     3     \\
15   &  33  &   4.4  $\pm$   0.9  &   25   &  57  &  0.32  $\pm$  0.07  &  34   &  70  &  0.16  $\pm$  0.03 &  43   &  95  &    10  $\pm$     2     \\
15   &  34  &   0.8  $\pm$   0.3  &   26   &  53  &   0.7  $\pm$   0.1  &  34   &  71  &   0.9  $\pm$   0.2 &  43   &  98  &   0.8  $\pm$   0.3     \\
15   &  35  &   0.4  $\pm$   0.1  &   26   &  54  &   3.2  $\pm$   0.6  &  34   &  72  &   3.5  $\pm$   0.7 &  44   &  93  &     5  $\pm$     2     \\
16   &  33  &     7  $\pm$     1  &   26   &  55  &     7  $\pm$     1  &  34   &  73  &     7  $\pm$     1 &  44   &  94  &     6  $\pm$     1     \\
16   &  34  &    12  $\pm$     2  &   26   &  56  &     8  $\pm$     2  &  34   &  74  &    10  $\pm$     2 &  44   &  95  &    15  $\pm$     3     \\
16   &  35  &   4.4  $\pm$   0.9  &   26   &  57  &   3.8  $\pm$   0.8  &  34   &  77  &   1.9  $\pm$   0.8 &  44   &  96  &    17  $\pm$     4     \\
16   &  36  &   1.1  $\pm$   0.2  &   26   &  58  &   1.3  $\pm$   0.3  &  34   &  78  &  0.23  $\pm$  0.08 &  44   &  97  &    11  $\pm$     2     \\
16   &  37  &  0.37  $\pm$  0.08  &   26   &  59  &  0.37  $\pm$  0.08  &  35   &  73  &  0.33  $\pm$  0.07 &  44   &  98  &     6  $\pm$     3     \\
17   &  35  &     6  $\pm$     1  &   27   &  55  &  0.40  $\pm$  0.08  &  35   &  74  &   1.4  $\pm$   0.3 &  45   &  95  &   2.2  $\pm$   0.9     \\
17   &  36  &     8  $\pm$     2  &   27   &  56  &   2.3  $\pm$   0.5  &  35   &  75  &     5  $\pm$     1 &  45   &  96  &   4.7  $\pm$   1.0     \\
17   &  37  &   4.9  $\pm$   1.0  &   27   &  57  &     6  $\pm$     1  &  35   &  76  &     9  $\pm$     2 &  45   &  97  &    14  $\pm$     3     \\
17   &  38  &   2.1  $\pm$   0.4  &   27   &  58  &     8  $\pm$     2  &  35   &  78  &   2.7  $\pm$   0.6 &  45   &  98  &    18  $\pm$     4     \\
17   &  39  &   0.5  $\pm$   0.1  &   27   &  59  &     6  $\pm$     1  &  35   &  80  &  0.28  $\pm$  0.06 &  45   &  99  &    15  $\pm$     3     \\
18   &  37  &   4.5  $\pm$   0.9  &   27   &  60  &   1.5  $\pm$   0.3  &  36   &  75  &   0.7  $\pm$   0.1 &  45   & 100  &     8  $\pm$     3     \\
18   &  38  &     8  $\pm$     2  &   27   &  61  &   0.5  $\pm$   0.1  &  36   &  76  &   3.2  $\pm$   0.7 &  46   &  97  &   1.1  $\pm$   0.4     \\
18   &  39  &     5  $\pm$     1  &   27   &  62  &  0.16  $\pm$  0.03  &  36   &  77  &     7  $\pm$     2 &  46   &  98  &   3.9  $\pm$   0.8     \\
18   &  40  &   2.5  $\pm$   0.5  &   28   &  57  &  0.24  $\pm$  0.05  &  36   &  78  &    12  $\pm$     2 &  46   &  99  &    12  $\pm$     3     \\
18   &  41  &   0.6  $\pm$   0.1  &   28   &  58  &   1.5  $\pm$   0.3  &  36   &  79  &     9  $\pm$     2 &  46   & 100  &    19  $\pm$     4     \\
18   &  42  &  0.22  $\pm$  0.05  &   28   &  59  &   4.9  $\pm$   1.0  &  36   &  80  &   3.1  $\pm$   0.7 &  46   & 101  &    19  $\pm$     4     \\
19   &  39  &   3.5  $\pm$   0.7  &   28   &  60  &     8  $\pm$     2  &  36   &  81  &   1.1  $\pm$   0.3 &  46   & 102  &    12  $\pm$     3     \\
19   &  40  &     7  $\pm$     1  &   28   &  61  &     7  $\pm$     1  &  37   &  78  &   2.4  $\pm$   0.5 &  47   & 100  &   2.8  $\pm$   0.6     \\
19   &  41  &     5  $\pm$     1  &   28   &  62  &   2.1  $\pm$   0.4  &  37   &  79  &     6  $\pm$     1 &  47   & 101  &    10  $\pm$     2     \\
19   &  42  &   2.8  $\pm$   0.6  &   28   &  63  &   0.7  $\pm$   0.1  &  37   &  80  &    11  $\pm$     2 &  47   & 102  &    17  $\pm$     3     \\
19   &  43  &   1.5  $\pm$   0.3  &   28   &  64  &  0.25  $\pm$  0.05  &  37   &  81  &     9  $\pm$     2 &  47   & 103  &    27  $\pm$     6     \\
19   &  44  &  0.29  $\pm$  0.06  &   29   &  60  &   1.0  $\pm$   0.2  &  37   &  82  &   3.8  $\pm$   0.8 &  48   & 102  &   1.6  $\pm$   0.3     \\
20   &  41  &   3.3  $\pm$   0.7  &   29   &  61  &   3.7  $\pm$   0.7  &  37   &  83  &   2.0  $\pm$   0.6 &  48   & 103  &     7  $\pm$     1     \\
20   &  42  &     7  $\pm$     1  &   29   &  62  &     7  $\pm$     1  &  37   &  84  &   1.5  $\pm$   0.4 &  48   & 104  &    16  $\pm$     3     \\
20   &  43  &     6  $\pm$     1  &   29   &  63  &     7  $\pm$     1  &  37   &  85  &  0.18  $\pm$  0.04 &  48   & 105  &    29  $\pm$     6     \\
20   &  44  &   3.6  $\pm$   0.7  &   29   &  64  &   2.6  $\pm$   0.5  &  38   &  80  &   1.8  $\pm$   0.4 &  48   & 106  &    24  $\pm$    10     \\
20   &  45  &   1.4  $\pm$   0.3  &   29   &  65  &   1.0  $\pm$   0.2  &  38   &  81  &     6  $\pm$     1 &  49   & 104  &   0.8  $\pm$   0.2     \\
20   &  46  &  0.34  $\pm$  0.07  &   29   &  66  &  0.27  $\pm$  0.06  &  38   &  82  &    12  $\pm$     2 &  49   & 105  &   3.5  $\pm$   0.7     \\
21   &  43  &   2.2  $\pm$   0.4  &   30   &  62  &   0.7  $\pm$   0.1  &  38   &  83  &    10  $\pm$     2 &  49   & 106  &     9  $\pm$     2     \\
21   &  44  &     6  $\pm$     1  &   30   &  63  &   3.1  $\pm$   0.6  &  38   &  84  &     6  $\pm$     1 &  49   & 107  &    24  $\pm$     5     \\
21   &  45  &     8  $\pm$     2  &   30   &  64  &     7  $\pm$     1  &  38   &  86  &   0.9  $\pm$   0.3 &  49   & 108  &    24  $\pm$     9     \\
21   &  46  &   3.6  $\pm$   0.7  &   30   &  65  &     8  $\pm$     2  &  38   &  87  &   0.5  $\pm$   0.1 &  50   & 106  &  0.32  $\pm$  0.07     \\
21   &  47  &   1.6  $\pm$   0.3  &   30   &  66  &   4.4  $\pm$   0.9  &  39   &  82  &   1.2  $\pm$   0.3 &  50   & 107  &   1.1  $\pm$   0.2     \\
21   &  48  &  0.40  $\pm$  0.08  &   30   &  67  &   1.3  $\pm$   0.3  &  39   &  83  &     6  $\pm$     2 &  50   & 108  &     9  $\pm$     2     \\
22   &  45  &   1.9  $\pm$   0.4  &   30   &  68  &   0.5  $\pm$   0.1  &  39   &  84  &    10  $\pm$     2 &  50   & 109  &    20  $\pm$     4     \\
22   &  46  &     6  $\pm$     1  &   30   &  69  &  0.18  $\pm$  0.04  &  39   &  85  &    14  $\pm$     3 &       &       &                       \\
22   &  47  &     8  $\pm$     2  &   31   &  64  &  0.43  $\pm$  0.09  &  39   &  86  &    20  $\pm$     6 &       &       &                       \\
22   &  48  &   4.7  $\pm$   0.9  &   31   &  65  &   2.0  $\pm$   0.4  &  39   &  88  &     5  $\pm$     2 &       &       &                       \\
\end{longtable*}

\begin{longtable*}{ll@{\hspace{5.5mm}}l@{\hspace{6.5mm}}ll@{\hspace{5.5mm}}l@{\hspace{6.5mm}}ll@{\hspace{5.5mm}}l@{\hspace{6.5mm}}ll@{\hspace{5.5mm}}l}
\caption[Production cross sections]{Production cross section of nuclides produced in reactions $^{124}$Sn + $^{124}$Sn at 1$\cdot A$ GeV. Data refers to the full production, corrected for the
limited angular acceptance of the FRS.} \label{124xstable} \\

\hline \hline \\[-2ex]
  \textbf{$Z$} & 
  \textbf{$A$} & 
  \textbf{$\sigma$} [mb]& 
  \textbf{$Z$} & 
  \textbf{$A$} & 
  \textbf{$\sigma$} [mb]&
  \textbf{$Z$} & 
  \textbf{$A$} & 
  \textbf{$\sigma$} [mb]&
  \textbf{$Z$} & 
  \textbf{$A$} & 
  \textbf{$\sigma$} [mb]\\[0.5ex] \hline
   \\[-1.8ex]
\endfirsthead

\multicolumn{12}{c}{{\tablename} \thetable{} -- Continued} \\[0.5ex]
  \hline \hline \\[-2ex]
  \textbf{$Z$} & 
  \textbf{$A$} & 
  \textbf{$\sigma$} [mb]& 
  \textbf{$Z$} & 
  \textbf{$A$} & 
  \textbf{$\sigma$} [mb]&
  \textbf{$Z$} & 
  \textbf{$A$} & 
  \textbf{$\sigma$} [mb]&
  \textbf{$Z$} & 
  \textbf{$A$} & 
  \textbf{$\sigma$} [mb]\\[0.5ex] \hline
  \\[-1.8ex]
\endhead


  \\[-1.8ex] \hline \hline
\endlastfoot
10   &  21  &    18  $\pm$     3 &  22   &  48  &     5  $\pm$     1 &  30   &  68  &   1.7  $\pm$   0.4  & 39   &  88  &     9  $\pm$     1   \\
10   &  22  &    19  $\pm$     4 &  22   &  49  &   3.3  $\pm$   0.7 &  30   &  69  &   1.0  $\pm$   0.2  & 39   &  89  &     4  $\pm$     1   \\
10   &  23  &     8  $\pm$     2 &  22   &  50  &   1.6  $\pm$   0.4 &  30   &  70  &  0.43  $\pm$  0.09  & 39   &  90  &   2.1  $\pm$   0.4   \\
11   &  23  &    18  $\pm$     3 &  22   &  51  &   0.6  $\pm$   0.1 &  30   &  71  &  0.14  $\pm$  0.06  & 39   &  91  &   1.1  $\pm$   0.2   \\
11   &  24  &    13  $\pm$     3 &  22   &  52  &  0.27  $\pm$  0.04 &  31   &  65  &   0.5  $\pm$   0.1  & 39   &  92  &  0.53  $\pm$  0.10   \\
11   &  25  &     9  $\pm$     2 &  23   &  47  &   0.5  $\pm$   0.1 &  31   &  66  &   1.6  $\pm$   0.4  & 39   &  93  &  0.23  $\pm$  0.04   \\
12   &  25  &    13  $\pm$     3 &  23   &  48  &   2.3  $\pm$   0.5 &  31   &  67  &   3.1  $\pm$   0.8  & 40   &  90  &     6  $\pm$     2   \\
12   &  26  &    15  $\pm$     3 &  23   &  49  &     5  $\pm$     1 &  31   &  68  &   3.5  $\pm$   0.9  & 40   &  91  &     5  $\pm$     2   \\
12   &  29  &   0.9  $\pm$   0.3 &  23   &  50  &     5  $\pm$     1 &  31   &  69  &   2.9  $\pm$   0.7  & 40   &  92  &   2.8  $\pm$   0.6   \\
13   &  27  &    12  $\pm$     3 &  23   &  51  &   3.6  $\pm$   0.8 &  31   &  70  &   2.0  $\pm$   0.4  & 40   &  93  &   1.7  $\pm$   0.3   \\
13   &  28  &    10  $\pm$     2 &  23   &  52  &   1.7  $\pm$   0.4 &  31   &  71  &   1.3  $\pm$   0.2  & 40   &  94  &   0.9  $\pm$   0.2   \\
13   &  29  &     7  $\pm$     1 &  23   &  53  &   0.9  $\pm$   0.2 &  31   &  72  &   0.7  $\pm$   0.1  & 40   &  95  &  0.44  $\pm$  0.08   \\
13   &  30  &   2.9  $\pm$   0.5 &  23   &  54  &  0.31  $\pm$  0.06 &  31   &  73  &  0.28  $\pm$  0.07  & 40   &  96  &  0.18  $\pm$  0.03   \\
14   &  29  &     9  $\pm$     2 &  24   &  49  &  0.36  $\pm$  0.08 &  32   &  67  &  0.33  $\pm$  0.07  & 41   &  92  &  10.4  $\pm$   0.3   \\
14   &  30  &    13  $\pm$     3 &  24   &  50  &   1.9  $\pm$   0.4 &  32   &  68  &   1.2  $\pm$   0.3  & 41   &  93  &     5  $\pm$     2   \\
14   &  31  &     6  $\pm$     1 &  24   &  51  &     4  $\pm$     1 &  32   &  69  &   2.5  $\pm$   0.7  & 41   &  94  &   3.6  $\pm$   1.0   \\
14   &  32  &   3.3  $\pm$   0.6 &  24   &  52  &     5  $\pm$     1 &  32   &  70  &   3.6  $\pm$   0.9  & 41   &  95  &   2.7  $\pm$   0.5   \\
14   &  33  &   1.4  $\pm$   0.2 &  24   &  53  &   3.5  $\pm$   0.8 &  32   &  71  &   2.9  $\pm$   0.7  & 41   &  96  &   1.5  $\pm$   0.3   \\
15   &  31  &     7  $\pm$     1 &  24   &  54  &   2.1  $\pm$   0.4 &  32   &  72  &   2.2  $\pm$   0.5  & 41   &  97  &   0.8  $\pm$   0.1   \\
15   &  32  &     8  $\pm$     2 &  24   &  55  &   0.9  $\pm$   0.2 &  32   &  73  &   1.5  $\pm$   0.2  & 41   &  98  &  0.34  $\pm$  0.06   \\
15   &  33  &     7  $\pm$     1 &  24   &  56  &  0.45  $\pm$  0.08 &  32   &  74  &   0.9  $\pm$   0.3  & 42   &  95  &     9  $\pm$     1   \\
15   &  34  &   3.2  $\pm$   0.6 &  25   &  51  &  0.25  $\pm$  0.06 &  32   &  75  &  0.42  $\pm$  0.09  & 42   &  96  &     5  $\pm$     1   \\
15   &  35  &   1.9  $\pm$   0.3 &  25   &  52  &   1.5  $\pm$   0.3 &  32   &  76  &  0.15  $\pm$  0.06  & 42   &  97  &   3.8  $\pm$   0.7   \\
15   &  36  &  0.69  $\pm$  0.08 &  25   &  53  &   3.9  $\pm$   0.9 &  33   &  69  &  0.19  $\pm$  0.04  & 42   &  98  &   2.4  $\pm$   0.4   \\
16   &  33  &     5  $\pm$     1 &  25   &  54  &     5  $\pm$     1 &  33   &  70  &   0.7  $\pm$   0.2  & 42   &  99  &   1.3  $\pm$   0.2   \\
16   &  34  &     9  $\pm$     2 &  25   &  55  &   3.9  $\pm$   0.9 &  33   &  71  &   1.8  $\pm$   0.5  & 42   & 100  &   0.7  $\pm$   0.1   \\
16   &  35  &     6  $\pm$     1 &  25   &  56  &   2.1  $\pm$   0.5 &  33   &  72  &   2.8  $\pm$   0.7  & 43   &  97  &     9  $\pm$     2   \\
16   &  36  &   3.6  $\pm$   0.7 &  25   &  57  &   1.2  $\pm$   0.3 &  33   &  73  &   3.1  $\pm$   0.7  & 43   &  98  &     6  $\pm$     3   \\
16   &  37  &   1.6  $\pm$   0.3 &  25   &  58  &   0.6  $\pm$   0.1 &  33   &  74  &   2.4  $\pm$   0.5  & 43   &  99  &     5  $\pm$     1   \\
16   &  38  &   0.8  $\pm$   0.1 &  26   &  54  &   1.1  $\pm$   0.3 &  33   &  75  &   1.8  $\pm$   0.3  & 43   & 100  &   3.6  $\pm$   0.6   \\
17   &  35  &   3.6  $\pm$   0.7 &  26   &  55  &   3.3  $\pm$   0.8 &  33   &  76  &   1.2  $\pm$   0.2  & 43   & 101  &   2.4  $\pm$   0.4   \\
17   &  36  &     6  $\pm$     1 &  26   &  56  &     5  $\pm$     1 &  33   &  77  &   0.6  $\pm$   0.1  & 43   & 102  &   1.3  $\pm$   0.3   \\
17   &  37  &     6  $\pm$     1 &  26   &  57  &   3.8  $\pm$   0.9 &  33   &  78  &  0.22  $\pm$  0.09  & 44   &  99  &     6  $\pm$     1   \\
17   &  38  &   3.5  $\pm$   0.7 &  26   &  58  &   2.6  $\pm$   0.6 &  34   &  71  &  0.14  $\pm$  0.03  & 44   & 100  &     8  $\pm$     1   \\
17   &  39  &   2.0  $\pm$   0.4 &  26   &  59  &   1.3  $\pm$   0.3 &  34   &  72  &   0.5  $\pm$   0.1  & 44   & 101  &     7  $\pm$     1   \\
17   &  40  &   1.0  $\pm$   0.2 &  26   &  60  &   0.7  $\pm$   0.2 &  34   &  73  &   1.2  $\pm$   0.3  & 44   & 102  &   5.5  $\pm$   0.9   \\
18   &  37  &   2.6  $\pm$   0.5 &  27   &  56  &   0.7  $\pm$   0.2 &  34   &  74  &   2.3  $\pm$   0.6  & 44   & 103  &   3.7  $\pm$   0.7   \\
18   &  38  &     6  $\pm$     1 &  27   &  57  &   2.6  $\pm$   0.6 &  34   &  75  &   2.6  $\pm$   0.6  & 44   & 104  &   2.3  $\pm$   0.4   \\
18   &  39  &     6  $\pm$     1 &  27   &  58  &     4  $\pm$     1 &  34   &  76  &   2.4  $\pm$   0.5  & 45   & 102  &    11  $\pm$     3   \\
18   &  40  &   4.0  $\pm$   0.8 &  27   &  59  &     4  $\pm$     1 &  34   &  77  &   1.9  $\pm$   0.4  & 45   & 103  &     8  $\pm$     2   \\
18   &  41  &   2.1  $\pm$   0.4 &  27   &  60  &   2.8  $\pm$   0.6 &  34   &  78  &   1.6  $\pm$   0.3  & 45   & 104  &     8  $\pm$     1   \\
18   &  42  &   1.1  $\pm$   0.2 &  27   &  61  &   1.7  $\pm$   0.4 &  34   &  79  &   0.9  $\pm$   0.2  & 45   & 105  &   5.9  $\pm$   0.8   \\
18   &  43  &  0.39  $\pm$  0.06 &  27   &  62  &   0.9  $\pm$   0.2 &  34   &  80  &  0.38  $\pm$  0.08  & 45   & 106  &   4.0  $\pm$   0.5   \\
19   &  39  &   1.9  $\pm$   0.4 &  27   &  63  &  0.42  $\pm$  0.10 &  35   &  74  &  0.15  $\pm$  0.06  & 46   & 104  &    15  $\pm$     6   \\
19   &  40  &     5  $\pm$     1 &  28   &  58  &  0.44  $\pm$  0.10 &  35   &  75  &   0.6  $\pm$   0.1  & 46   & 105  &    10  $\pm$     4   \\
19   &  41  &     6  $\pm$     1 &  28   &  59  &   1.8  $\pm$   0.4 &  35   &  76  &   1.3  $\pm$   0.3  & 46   & 106  &    11  $\pm$     2   \\
19   &  42  &   4.2  $\pm$   0.8 &  28   &  60  &   3.8  $\pm$   0.9 &  35   &  77  &   2.0  $\pm$   0.5  & 46   & 107  &     8  $\pm$     1   \\
19   &  43  &   2.6  $\pm$   0.5 &  28   &  61  &     4  $\pm$     1 &  35   &  78  &   1.9  $\pm$   0.4  & 46   & 108  &     7  $\pm$     1   \\
19   &  44  &   1.3  $\pm$   0.2 &  28   &  62  &   3.4  $\pm$   0.8 &  35   &  79  &   1.9  $\pm$   0.5  & 47   & 106  &    10  $\pm$     3   \\
19   &  45  &   0.6  $\pm$   0.1 &  28   &  63  &   2.0  $\pm$   0.4 &  35   &  80  &   1.8  $\pm$   0.4  & 47   & 107  &    11  $\pm$     2   \\
20   &  41  &   1.7  $\pm$   0.4 &  28   &  64  &   1.1  $\pm$   0.2 &  35   &  81  &   1.2  $\pm$   0.2  & 47   & 108  &    12  $\pm$     2   \\
20   &  42  &   4.5  $\pm$   1.0 &  28   &  65  &   0.6  $\pm$   0.1 &  35   &  82  &   0.5  $\pm$   0.1  & 47   & 109  &    12  $\pm$     1   \\
20   &  43  &     6  $\pm$     1 &  28   &  66  &  0.20  $\pm$  0.04 &  36   &  81  &     5  $\pm$     1  & 47   & 110  &    11  $\pm$     3   \\
20   &  44  &     5  $\pm$     1 &  29   &  60  &  0.22  $\pm$  0.05 &  36   &  82  &     3  $\pm$     1  & 48   & 109  &    12  $\pm$     5   \\
20   &  45  &   2.6  $\pm$   0.6 &  29   &  61  &   1.2  $\pm$   0.3 &  36   &  83  &   1.8  $\pm$   0.3  & 48   & 110  &    12  $\pm$     4   \\
20   &  46  &   1.5  $\pm$   0.3 &  29   &  62  &   2.8  $\pm$   0.7 &  36   &  84  &   0.9  $\pm$   0.2  & 48   & 111  &    15  $\pm$     6   \\
20   &  47  &  0.59  $\pm$  0.08 &  29   &  63  &     4  $\pm$     1 &  36   &  85  &  0.36  $\pm$  0.06  & 48   & 112  &    15  $\pm$     6   \\
21   &  43  &   1.0  $\pm$   0.2 &  29   &  64  &   3.4  $\pm$   0.8 &  37   &  84  &     4  $\pm$     1  & 49   & 111  &    12  $\pm$     3   \\
21   &  44  &   3.4  $\pm$   0.7 &  29   &  65  &   2.4  $\pm$   0.5 &  37   &  85  &   2.4  $\pm$   0.5  & 49   & 112  &    10  $\pm$     2   \\
21   &  45  &     6  $\pm$     1 &  29   &  66  &   1.4  $\pm$   0.3 &  37   &  86  &   1.2  $\pm$   0.2  & 49   & 113  &    16  $\pm$     2   \\
21   &  46  &   4.7  $\pm$   1.0 &  29   &  67  &   0.8  $\pm$   0.2 &  37   &  87  &  0.51  $\pm$  0.09  & 49   & 114  &    15  $\pm$     3   \\
21   &  47  &   3.0  $\pm$   0.6 &  29   &  68  &  0.31  $\pm$  0.07 &  37   &  88  &  0.19  $\pm$  0.04  & 50   & 113  &     6  $\pm$     2   \\
21   &  48  &   1.6  $\pm$   0.3 &  30   &  63  &   0.9  $\pm$   0.2 &  38   &  86  &     5  $\pm$     1  & 50   & 114  &     6  $\pm$     1   \\
21   &  49  &   0.6  $\pm$   0.1 &  30   &  64  &   2.5  $\pm$   0.6 &  38   &  87  &   3.0  $\pm$   0.7  & 50   & 115  &    11  $\pm$     2   \\
22   &  45  &   0.7  $\pm$   0.2 &  30   &  65  &   3.8  $\pm$   1.0 &  38   &  88  &   1.5  $\pm$   0.3  & 50   & 116  &    12  $\pm$     5   \\
22   &  46  &   3.2  $\pm$   0.7 &  30   &  66  &   3.8  $\pm$   1.0 &  38   &  89  &   0.8  $\pm$   0.1  &      &      &                      \\
22   &  47  &     5  $\pm$     1 &  30   &  67  &   2.7  $\pm$   0.6 &  38   &  90  &  0.32  $\pm$  0.06  &      &      &                      \\
\end{longtable*}

\bibliographystyle{h-physrev}
\bibliography{references}
\end{document}